\documentclass[aps,prx,twocolumn,groupedaddress,floatfix]{revtex4-2}
\usepackage{amsmath}
\usepackage{amsfonts}
\usepackage{physics}
\usepackage{graphicx}
\usepackage{float}
\usepackage{array}
\usepackage{soul}
\usepackage{color}
\usepackage{verbatim}
\usepackage{subfigure}

\begin{document}
    \title{Detection of anyon braiding through pump-probe spectroscopy}
    
    \author{Xu Yang}
    \email{xu.yang@ntu.edu.sg}
    \affiliation{Department of Physics, The Ohio State University, 191 West Woodruff Avenue, Columbus, Ohio  43210-1117}
    \affiliation{School of Physical and Mathematical Sciences, Nanyang Technological University,SPMS-04-01,
21 Nanyang Link,
Singapore 637371}

    \author{Ryan Buechele}
    \email{buechele.5@osu.edu}
    \affiliation{Department of Physics, The Ohio State University, 191 West Woodruff Avenue, Columbus, Ohio  43210-1117}

    \author{Nandini Trivedi}
    \email{trivedi.15@osu.edu}
    \affiliation{Department of Physics, The Ohio State University, 191 West Woodruff Avenue, Columbus, Ohio  43210-1117}

    \date{\today}

\begin{abstract}
    
    We show that the braiding of anyons in a quantum spin liquid leaves a distinct dynamical signature in the nonlinear pump-probe response. Using a combination of exact diagonalization and matrix product state techniques, we study the nonlinear pump-probe response of the toric code in a magnetic field, a model which hosts mobile electric $e$ and magnetic $m$ anyonic excitations. While the linear response signal oscillates and decays with time like $\sim t^{-1.3}$, the amplitude of the nonlinear signal for $\chi^{(3)}_{XZZ}$ features a linear-in-time enhancement at early times. The comparison between $\chi^{(3)}_{XZZ}$, which involves the non-trivial braiding of $e$ and $m$ anyons, and $\chi^{(3)}_{XXX}$ that involves the trivial braiding of the same types of anyons,
    serves to distinguish the braiding statistics of anyons. We support our analysis by constructing a hard-core anyon model with statistical gauge fields to develop further insights into the time dependence of the pump-probe response. Pump-probe spectroscopy provides a distinctive new probe of quantum spin liquid states, beyond the inconclusive broad features observed in single spin flip inelastic neutron scattering. 

\medskip


\end{abstract}

\maketitle

\section{Introduction}
Anyons are exotic quasiparticle (QP) excitations\cite{leinaas1977theory,wilczek1990fractional,frohlich1988quantum,frohlich1989quantum}, that can emerge in topologically-ordered two-dimensional systems such as fractional quantum Hall (FQH) states\cite{tsui1982two,laughlin1983anomalous,cage2012quantum} and quantum spin liquids (QSLs)\cite{anderson1987resonating,lee2006doping,savary2016quantum,zhou2017quantum}. These anyonic excitations carry only a fraction of the original quantum numbers of the system, and exhibit nontrivial braiding statistics\cite{barkeshli2019symmetry}, meaning that one anyon encircling another alters the total wave function of the system. Experimentally identifying anyons in a material is a necessary step in realizing fault-tolerant quantum computation 
%and understanding certain high-temperature superconductors
\cite{nayak_2008_review}, but also presents a variety of challenges. The fractionalization of QPs appears in inelastic neutron scattering data as a broad continuum in the spectrum \cite{mourigal_spinonINS_2013, han_2012_KagomeINS, banerjee_2017_RuCl3INS} due to the many ways that fractionalized QPs can share momentum and energy, but this is often hard to distinguish from other sources of broadening in the measurement, such as disorder. Recently, some studies have suggested examining the spectrum of two-spin excitations in the ground state for sharp features of fractionalized phases \cite{feng_anyon_2023}. These multi-spin correlators within linear response are currently being explored by noise spectroscopy~\cite{takahashi2025spiral, maze2008nanoscale}. 

Observing the effects of braiding in bulk QSLs, however, requires a new approach to identify a unique characteristic of anyon braiding in an experimentally measurable quantity. 
%Most studies have focused on measuring the spin structure factor available from linear response measurements, which fails to capture the physics of braiding. 
Braiding statistics have only recently been observed in fractional quantum Hall states, whose chiral edge modes allow for careful control of anyon trajectories in order to interfere states with different braiding-induced phases \cite{bartolomei_2020_anyon-statistics, nakamura_2020_anyon-braiding}. We propose using nonlinear pump-probe spectroscopy to detect, in addition to fractionalization, the braiding of anyons around each other by using magnetic fields that couple to non-commuting operators. The resulting nonlinear response of the multi-time correlation functions allows us to uncover experimental evidence of braiding. 

\begin{figure}[t]
    \centering
    \includegraphics[width=\linewidth]{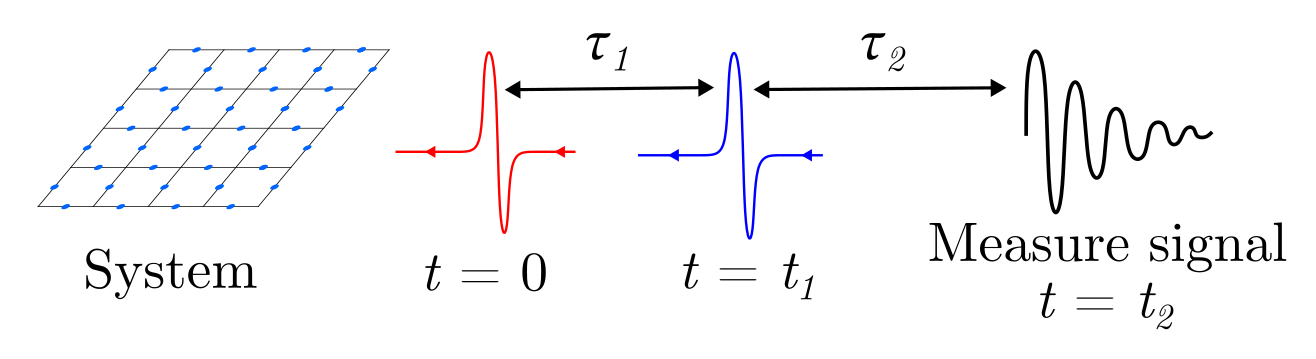}
    \caption{\textbf{Experimental Schematic:} Diagram of the pump-probe experimental process. A first ``pump'' pulse is applied to the system at $t=0$, then a second ``probe'' pulse is applied at $t=t_1$. The resulting signal output is measured at time $t=t_2$.}
    \label{fig:exp-schematic}
\end{figure}

Probing pump-probe signals with 2D coherent spectroscopy in the frequency domain has been used to identify ``hot spots'' that reflect energies of excitations in quantum materials \cite{hebling2008high,lu2017coherent,armitage_2dcs_2019,sim2023nonlinear,sim2023microscopic,choi_theory_2020,gao2023two,nandkishore2021spectroscopic,hart2023extracting,qiang2024probing,watanabe2024exploring,watanabe2024revealing,zhang2024disentangling}. More recent proposals have focused on the long-time amplitude of signals of the nonlinear response in the real-time domain \cite{morampudi2017statistics,fava_divergent_2023,mcginley_pump_probe_prl}. We build on the latter approach with new results in the early to intermediate time regime that we claim is crucial to see the signatures of braiding between anyons. We consider a pump-probe process where a system is perturbed from equilibrium by two pulses with a variable delay between them, as depicted schematically in Fig. \ref{fig:exp-schematic}. The QPs excited by these pulses have the ability to propagate and braid with each other, and the accumulation of statistical braiding phase in the wave function leads to a distinct early-time behavior of the measured response function.

A large portion of prior unbiased numerical studies such as those based on density matrix renormalization group (DMRG) and exact diagonalization (ED) are mostly restricted to one-dimensional systems and primarily concern the energy of quasiparticle excitations \cite{sim2023nonlinear,sim2023microscopic,fava_divergent_2023,hart2023extracting,li2023photon,watanabe2024revealing,watanabe2024exploring}. Our work focuses on two-dimensional (2D) systems to discuss the signatures of anyon braiding statistics in nonlinear response; in particular we consider the pump-probe responses of the toric code in an applied magnetic field, a model with a simple effective description in terms of dynamic anyons. We present below the first study to employ infinite DMRG \cite{white1992density,schollwock2011density,mcculloch2008infinite} together with time-dependent variational principle (TDVP)\cite{paeckel2019time, haegeman2011time, haegeman2013post, haegeman2016unifying, yang2020time} for nonlinear response analysis in 2D systems, supplemented by ED \cite{sandvik_2010,lauchli2010numerical} to corroborate our iDMRG results at short times. We gain further analytical insight by using a large-scale anyon-hopping model \cite{hatsugai1991anyons,hatsugai1991braid,kirchner2023numerical} to systematically investigate how the signatures of anyon statistics manifest in nonlinear responses. This approach clearly distinguishes statistical effects from conventional non-statistical interactions, providing a quantitative framework for anyon dynamics to guide numerical simulations and experiments. We show in Fig. \ref{fig:chi3-plots} how the third-order pump-probe response features a linear-in-time growth relative to linear response only when anyon braiding occurs in the system, distinct from the signature in cases without braiding; furthermore, we explicitly reproduce this early time behavior in the anyon-hopping model along with analytic forms for the nonlinear response in Fig. \ref{fig:chi3-fitting}. Our complete results pave the way for new insights into the topological nature of these anyonic excitations and guides the experimental detection of anyons in quantum materials and cold atom platforms.

\section{Model: Toric Code in Magnetic Field}

\begin{figure*}
    \centering
    \includegraphics[width=0.9\textwidth]{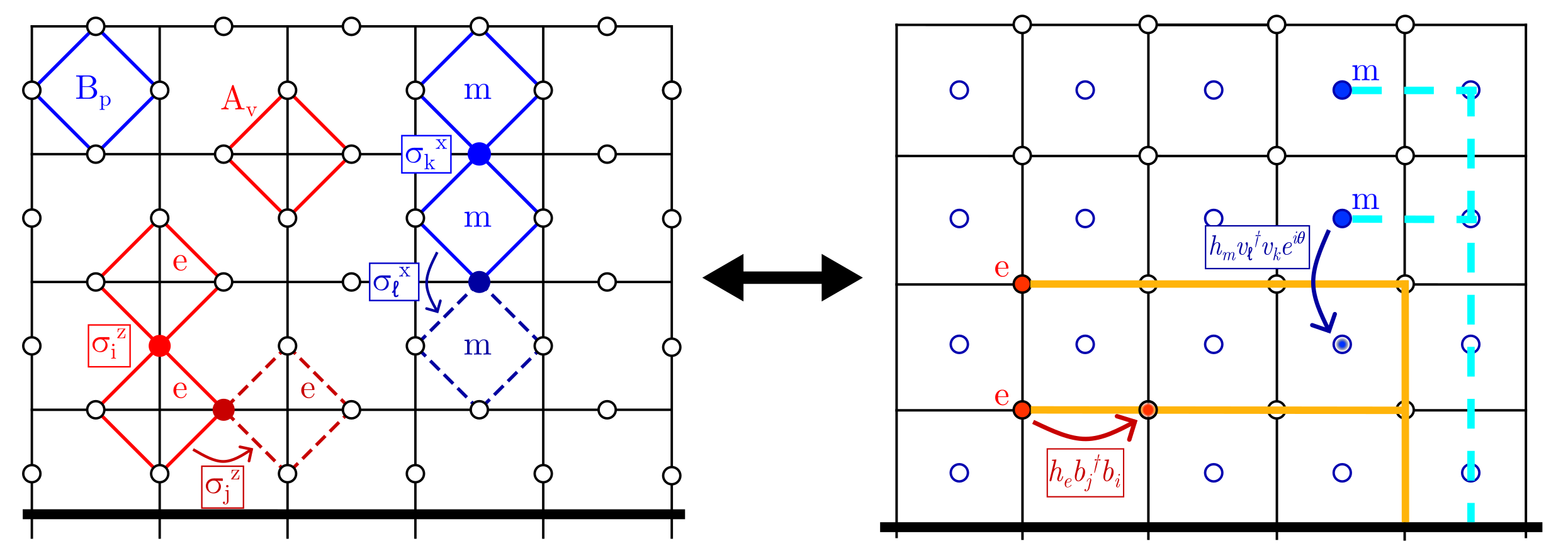}
    \caption{\textbf{Toric Code Lattice:}
    On the left, the toric code with spin-1/2 sites located on the links of a 2D square lattice. The star operator $A_v$ is the product of $\sigma^x$ on the sites neighboring a vertex $v$, while the plaquette operator $B_p$ is the product of $\sigma^z$ on the sites neighboring a plaquette $p$. $\sigma^z$ at a site $i$ flips both adjacent stars, creating two $e$ excitations, which can then hop when $\sigma^z$ acts on a neighboring site $j$. Likewise, $\sigma^x$ at site $k$ flips both adjacent plaquettes and creates two $m$ excitations, which can hop by the action of $\sigma^x$ on a neighboring site $\ell$. We compute the response functions on a cylindrical geometry; the thick line at the bottom represents the periodic boundary. 
    On the right, we model the toric code excitations as hard-core anyons hopping on a square lattice and its dual lattice; the $e$'s live on the lattice sites and are created/annihilated by the operators $b,b^\dagger$ at each site, while the $m$'s live on the dual lattice and are created/annihilated by the operators $v, v^\dagger$. Each anyon carries with it a string (solid orange for $e$, dashed blue for $m$) which runs to the rightward then downward to terminate at the periodic boundary. Tracking how these strings sweep over each other accounts for the phase $\theta_{em}$ acquired when $e$'s encircle an $m$ (or vice versa), as well as any phase acquired from incontractible loops around the periodic system \cite{hatsugai1991anyons}.}
    \label{fig:tc-lattice-diagram}
\end{figure*}

To demonstrate the principles behind anyon detection using signatures from nonlinear response, we consider the toric code with spin-1/2 qubits $\sigma=(\sigma^x,\sigma^y,\sigma^z)$ on the links of a square lattice in an external magnetic field given by:
\begin{equation}
    \mathcal{H}_{TC} = -J_A \sum_v A_v - J_B \sum_p B_p - h_x\sum_j \sigma_j^x -h_z\sum_j \sigma_j^z
    \label{eq:TC_Ham}
\end{equation}
where the star operator $A_v = \prod_{j\in v} \sigma_j^x$ is a product of $\sigma_x$ on sites adjacent to the vertex $v$ and $B_p = \prod_{j\in p} \sigma_j^z$ is a product of $\sigma_z$ on sites surrounding a plaquette $p$, as shown in Fig. \ref{fig:tc-lattice-diagram}. Throughout, we assume $J_A = J_B \equiv J$ and $h_x=h_x\equiv h$ unless otherwise noted. In the absence of the field terms ($h=0$), the ground state is determined by the spin configuration which makes $A_v = 1, B_p=1\ \forall v,p$ \cite{kitaev_fault-tolerant_2003}. Gapped excitations above the ground state are created by spin-flips which fractionalize into two anyons; these QPs can only be created/annihilated in pairs and each pair costs an energy $4J$. Applying $\sigma^z$ to a site flips the $x$-projection of the spin, which flips the sign of both adjacent star operators and creates two stationary $e$ particles; $\sigma^x$ on a site flips the $z$-projection of spin, flipping both adjacent plaquette operators and creating two stationary $m$ particles. Turning on a perturbatively small field in the $z(x)$-direction allows the $e$($m$) QPs to hop around the lattice. The $e, m$ particles are individually bosons but have mutual semionic statistics, meaning that braiding an $e$ around an $m$ contributes a phase factor of $-1$ to the wave function. This effective description of $e,m$ excitations hopping holds for $h \lesssim 0.3 J$; for larger values of the external field, the model undergoes a phase transition to a polarized state along the direction of the applied field \cite{vidal_low-energy_2009,wu_phase_2012}. In the topologically trivial field-polarized phase with $h \gg J$, the low-energy excitations are single spin flips which can hop around the lattice via the star and plaquette operators but have trivial braiding statistics.

We use three complementary numerical tools to study the linear and nonlinear responses:
\begin{enumerate}
    \item Exact diagonalization method. ED is constrained by the exponentially large Hilbert space dimension, and practically the linear dimension $L$ can be as large as $\sim 4-6$ depending on the specific lattice geometry. We choose a $L_x\times L_y$ lattice with $L_x=3,L_y=4$ with periodic boundary conditions in our study.
    \item iDMRG method combined with TDVP. The iDMRG is done on a $L_x\times L_y$ cylinder with $L_x$ infinite and $L_y$ finite and periodic. The computational complexity scales exponentially with $L_y$, constraining the practically achievable $L_y$ to consist of unit-cells as many as $5-6$. We choose to study systems with $L_y=3$ throughout this work.
    \item Anyon hopping model. We have also used an anyon-hopping model with hard-core anyons hopping on a $L_x\times L_y$ square lattice to simulate the anyon dynamics. The Hamiltonian is obtained by projecting the full many-body Hilbert space onto the subspace with fixed number of hard-core anyons which captures the essential physics of anyon braiding statistics. The accessible system size can be as large as $L\sim 20$ or beyond, far exceeding the reach of ED and iDMRG.
\end{enumerate}

The details of the implementation of these numerical methods are given in Section \ref{section:methods}. Before concluding this section, we briefly comment on the role of Lieb-Robinson velocity $v_{LR}$ and the finite-size effect. Using a Lieb-Robinson bound \cite{lieb1972finite}, the evolution of a local operator after time $t$ only affects operators within a light cone of radius $v_{LR}t$; therefore, the finite-size effects becomes apparent after time $t\sim L/(2v_{LR})$. This sets the upper bound on the time scale that are free from finite-size effects in numerical simulations. The Lieb-Robinson velocity is lower-bounded by the maximum velocity of quasiparticles. We estimate $v_{LR}\sim 2h_{\text{kinetic}}$ (where $h_{\text{kinetic}}$ denotes the hopping strength of relevant anyons-in our case, $h_{x/z}$). Therefore, we can use the time scale $L/(2v_{LR})\sim L/(4h_{\text{kinetic}})$ to estimate the time range that faithfully reflects the thermodynamic limit. 

\section{Linear Response}

\begin{figure}
    \centering
    \includegraphics[width=0.88\linewidth]{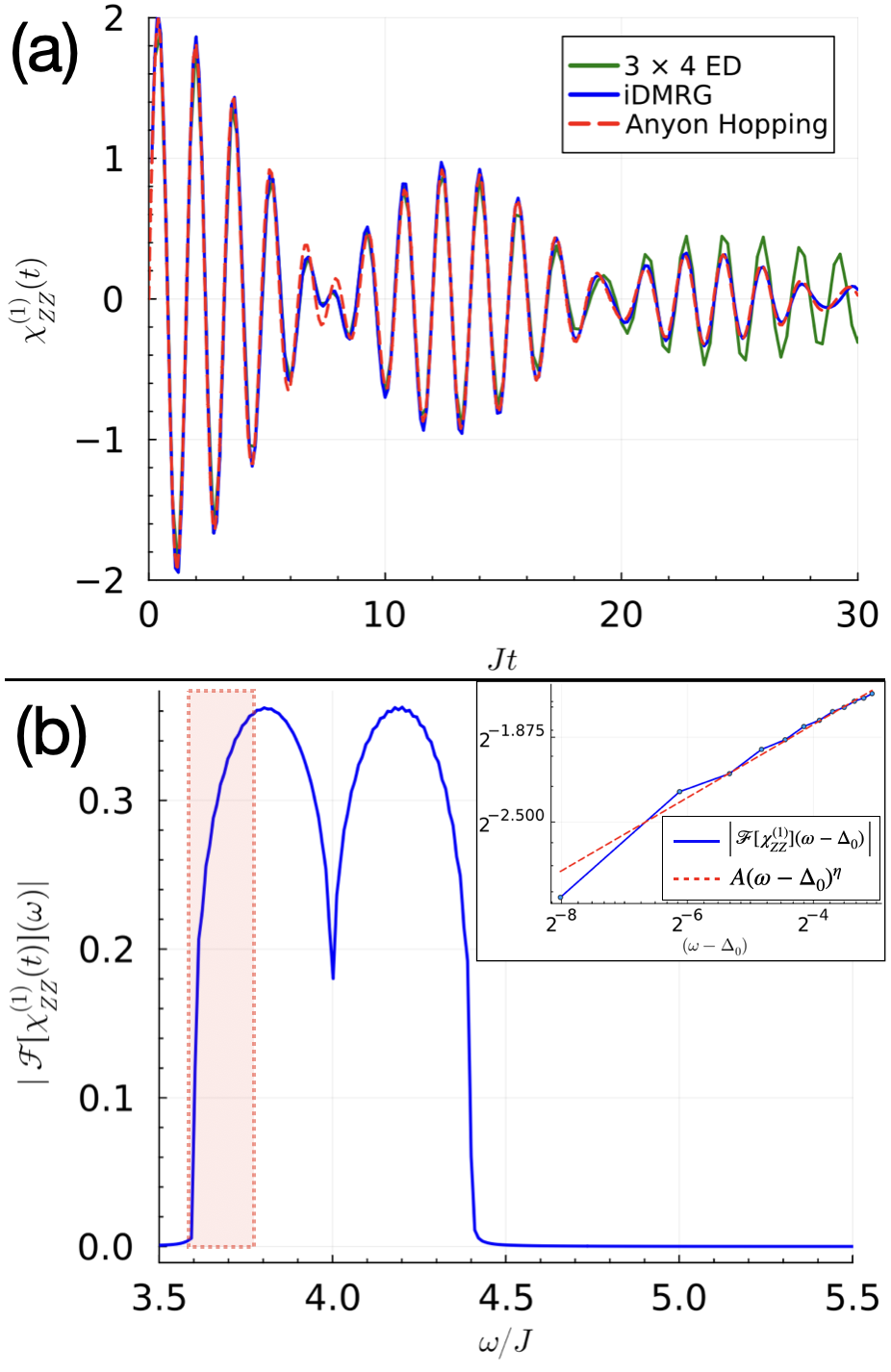}
    \caption{\textbf{Linear Response:}
    \textbf{(a)} Linear Response $\chi_{ZZ}^{(1)}(t)$ computed with ED, iDMRG, and the anyon-hopping model with $J=1, h=0.05$.  $\chi^{(1)}_{ZZ}(t)$ characterizes the probability that a pair of $e$ particles annihilate after time $t$; therefore, in finite systems, the probability of meeting at late time is enhanced, causing the ED data to deviate from the iDMRG data where $L_x\rightarrow \infty$. In the anyon-hopping model we have used the un-renormalized parameters $h_e=0.05,\Delta_e=2$. The agreement between the iDMRG and anyon-hopping model is excellent. The deviation of ED data from the iDMRG data after $Jt\sim 16$ is due to the finite size in $L_x$.
    \textbf{(b)} Magnitude of the Fourier transform of $\chi_{ZZ}^{(1)}(t)$, computed with the hard-core anyon-hopping model on a $100\times 100$ lattice with periodic boundary conditions up to $Jt=300$; the large system was used to minimize finite size effects at the long times needed for clear momentum resolution. This matches the expected behavior for hard-core bosons \cite{morampudi2017statistics}. \textit{Inset:} To illustrate the power-law dependence of the Fourier transformed signal at the onset, we have shown on a log-log plot just the highlighted data immediately above the gap energy $\Delta_0$ fit to a form $A(\omega-\Delta_0)^\eta$ and find $\eta\approx0.27$, suggesting the real time signal decays like $1/t^{-(1+\eta)}$. 
    }  
    \label{fig:linear_plots}
\end{figure}

In linear response, the system in its ground state is first perturbed by a pulse considered here to couple to the total magnetization operator $M^\alpha \equiv \sum_{j=1}^N \sigma_j^\alpha \ (\alpha\in\{x,y,z\})$ at $t=0$. The resulting magnetization $M^\beta$ is then measured at a later time $t$ from which we extract the linear response coefficient \cite{mukamel_principles_1999, hamm_principles_2005}
\begin{equation}
    \chi_{\alpha,\beta}^{(1)}(t) = -\frac{i}{N} \langle 0| [M^\alpha(0), M^\beta(t)] |0\rangle
    \label{eq:lin_definition}
\end{equation}
where $M^\alpha(t) \equiv e^{i\mathcal{H}t} M^\alpha e^{-i\mathcal{H}t}$ is the time-evolved operator $M^\alpha$ in the Heisenberg picture and $\ket{0}$ denotes the ground state of the system. Our results for $\chi^{(1)}_{ZZ}(t)$ computed with both ED and iDMRG methods are shown in Fig. \ref{fig:linear_plots}(a); the ED calculations are done using Krylov subspace expansions to compute the matrix exponentiation for time evolution \cite{Haegeman_KrylovKit_2024}. Specifically for the $\mathbb{Z}_2$ spin liquid phase of the toric code model, the pulse that couples to $M^z(0)$ excites a pair of $e$ QPs with a gap $2J$ (per QP), and these $e$'s can propagate through the system when there is a nonzero $h_z$ before being annihilated by $M^z(t)$. We therefore expect the linear response to feature an oscillation at a frequency equal to the energy gap $4J$ to create the anyon pair, with a decaying envelope related to the fact that excited anyons must form a closed space-time loop to annihilate.

For a pair of free bosons, the decaying envelope for the linear response is predicted to scale as $1/t$ \cite{mcginley_pump_probe_prl}; however, numerically we find that the long-time behavior of the pump-probe response as computed from iDMRG/ED decays faster. This is evident in the Fourier Transform of $\chi^{(1)}_{ZZ}$, shown in Fig. \ref{fig:linear_plots}(b); the Fourier transform of $1/t$ should give a step function at the onset gap $\Delta_0$, but we instead see a power law growth $\propto (\omega-\Delta_0)^\eta$ which corresponds to a decay like $t^{-(1+\eta)}$ (for $\eta>0$). We argue this faster decay results from the hard-core constraint between anyons that serves as an effective short-range repulsion and decreases the probability of two anyons meeting to annihilate at time $t$ \cite{morampudi2017statistics}.  
        
In order to better capture the effect of hard-core constraints in anyons and fit the numerical data, we have also computed the linear response based on a simple anyon-hopping model \cite{hatsugai1991braid,hatsugai1991anyons,kirchner2023numerical} with hard-core anyons hopping on a $L_x\times L_y$ square lattice, with Hamiltonian given by
\begin{align}
    \mathcal{H}_{e,m}=&-h_e\sum\limits_{<jk>}b_j^{\dagger}b_ke^{i\theta^e_{jk}}-h_m\sum\limits_{<jk>}v_j^{\dagger}v_ke^{i\theta^m_{jk}} \nonumber \\ 
    &+\Delta_e N_e+\Delta_m N_m,
    \label{eq:anyon_hopping_ham}
\end{align} 
where $b^{\dagger}/v^{\dagger}$ are the creation operators of the $e/m$ particle, and each anyon carries a string which encodes the phase angle that tracks braiding phases $\theta^{e/m}$ as delineated in detail in Refs.\cite{hatsugai1991braid,hatsugai1991anyons,kirchner2023numerical}. The operators $N_{e/m}$ count the total number of $e/m$ particles in the Hilbert space. The hopping parameters $h_{e/m}$ and gaps $\Delta_{e/m}$ directly correspond to the external field and excitation gap (per particle) in $\mathcal{H}_{TC}$; thus we choose the parameters $h_e=h_m=h$ and $\Delta_e=\Delta_m=2J$. For $\chi^{(1)}_{ZZ}$, we enforce hard-core constraints and restrict the Hamiltonian Eq. \eqref{eq:anyon_hopping_ham} to a subspace with exactly 2 $e$'s. This restriction is justified by the large energy penalty $\Delta_{e/m}$ for changing anyon numbers, ensuring the model remains valid up to timescales where non-statistical interactions become significant. 

Discarding the hard-core constraint for the moment, we can diagonalize the anyon-hopping Hamiltonian and obtain the dispersion
\begin{equation}
    \epsilon_{\textbf{k}}=-2h_e(\text{cos}(k_x)+\text{cos}(k_y))+\Delta_e.
    \label{eq:anyon_dispersion}
\end{equation}
The linear response can then be evaluated exactly as 
\begin{equation}
    \chi^{(1)}_{ZZ,\mathrm{free}}(t) = 2\sin(2\Delta_e t)J_0^2(4h_et),
    \label{eq:free-boson-chi1}
\end{equation}
where $J_0(x)$ is the zeroth-order Bessel function of the first kind. For $t\gg 1/h_e$, the Bessel function can be asymptotically approximated by
\begin{align}\label{eq:1/t}
   & J_0(4h_et)^2\approx \frac{\text{cos}^2(4h_e t-\pi/4)}{2\pi h_e t}
\end{align}
which leads to the expected $1/t$ decay for a pair of non-interacting bosonic anyons\cite{morampudi2017statistics,mcginley2024anomalous,mcginley_pump_probe_prl}. Including hard-core constraints significantly change the late-time behavior. Results of a large-scale simulation of hard-core anyon-hopping model on a $100\times 100$ lattice are shown in Fig. \ref{fig:linear_plots}; fitting the Fourier transformed data immediately above the gap $\Delta_0\equiv2(\Delta_e-4h_e)$ to a power law like $A(\omega-\Delta_0)^\eta$ yields $\eta\approx0.27$, confirming that hard-core constraints lead to a faster long-time decay like $t^{-(1+\eta)}$.

\section{Nonlinear Response}
    
\noindent {\bf {Formalism:}} We next consider a two-pulse experiment where the system is first excited by a ``pump'' pulse at $t=0$ coupled to $M^\alpha$; we assume the pulse has a short enough duration that it effectively instantly evolves the system from the ground state $\ket{0}$ to the state $e^{-i\mu M^\alpha}\ket{0}$, where $\mu$ is the coupling strength to the pump pulse. Then a second ``probe'' pulse is sent at $t=t_1$ coupled to $M^\beta$, before finally measuring the resulting signal $M^\gamma$ at a time $t_2$. (Throughout, we use the symbol $t_j$ to indicate an absolute time, while the Greek $\tau_j \equiv t_j - t_{j-1}$ indicates a time interval between subsequent times.) This experiment is repeated without the pump pulse, and the second signal is subtracted from the first to extract the pump-probe (PP) signal. This subtraction removes any contributions to the signal from linear response, leaving only higher-order contributions. 

The pump-probe coefficient we study is defined as\cite{mukamel_principles_1999, armitage_2dcs_2019}
\begin{align}
    \Xi_{\alpha, \beta, \gamma}^{PP}(t_1,& t_2) = \nonumber \\ 
    &-\frac{i}{N} \Big( \underset{\chi^{(1)} \text{in pumped state}}{\underbrace{\langle 0| e^{i \mu M^\alpha} [M^\beta(t_1), M^\gamma(t_2)] e^{-i \mu M^\alpha} |0\rangle}} \nonumber \\ 
    &- \underset{\chi^{(1)}\text{in g.s.}}{\underbrace{\langle 0| [M^\beta(t_1), M^\gamma(t_2)] |0\rangle}} \Big);
    \label{eq:XiPP_definition}
\end{align}
we expand the state $e^{-i \mu M^\alpha(0)} \ket{0}$ to second-order in the limit of a weak pulse $\mu$
\begin{align}
    e&^{-i \mu M^\alpha(0)} \ket{0} \nonumber \\
    &= \left(\mathbb{I} - i\mu M^{\alpha}(0) -\frac{\mu^2}{2} (M^{\alpha}(0))^2  + \mathcal{O}(\mu^3)\right) \ket{0}.
\end{align}
Substituting this into the first term of Eq. \eqref{eq:XiPP_definition}, the unperturbed linear response vanishes, and we collect the remaining terms in powers of $\mu$ to obtain 
\begin{align}
    \Xi_{\alpha, \beta, \gamma}^{PP}&(t_1, t_2) = \frac{\mu}{N} \bra{0} [M^\alpha(0), [M^\beta(t_1), M^\gamma(t_2)]] \ket{0} \nonumber \\
    &+ \frac{i\mu^2}{2N}\bra{0} [M^\alpha(0), [M^\alpha(0), [M^\beta(t_1), M^\gamma(t_2)]]] \ket{0} \nonumber \\
    &+ \mathcal{O}(\mu^3).
\end{align}
We can identify these terms as the second- and third-order response coefficients, following conventions from nonlinear response theory (see Appendix \ref{appendix:nonlinear_review} for more details), and expand the commutators to obtain:
\begin{align}
    \chi_{\alpha, \beta, \gamma}^{(2)}(t_1,& t_2) 
    = \nonumber \\ 
    &\frac{2}{N} \Re\Big[\bra{0}M^\alpha(0)M^\gamma(t_2)M^\beta(t_1)\ket{0} \nonumber \\ 
    &- \bra{0}M^\alpha(0)M^\beta(t_1)M^\gamma(t_2)\ket{0} \Big]
    \label{eq:chi2_definition}
\end{align}
\begin{align}
    \chi_{\alpha, \beta, \gamma}^{(3)}(t_1,& t_2)
    = \nonumber \\ 
    &-\frac{2}{N} \Im[2\bra{0} M^\alpha(0)M^\beta(t_1)M^\gamma(t_2)M^\alpha(0) \ket{0}  \nonumber \\
    &+ \bra{0} M^\alpha(0)M^\alpha(0)M^\gamma(t_2)M^\beta(t_1) \ket{0} \nonumber \\ 
    &+ \bra{0} M^\gamma(t_2)M^\beta(t_1)M^\alpha(0)M^\alpha(0) \ket{0} ].
    \label{eq:chi3_definition}
\end{align}
\begin{equation}
    \Xi_{\alpha, \beta, \gamma}^{PP}(t_1, t_2) = \mu \chi^{(2)}_{\alpha,\beta,\gamma}(t_1,t_2) + \frac{\mu^2}{2} \chi^{(3)}_{\alpha,\beta,\gamma}(t_1,t_2) + \mathcal{O}(\mu^3)
    \label{eq:Xi_expanded}
\end{equation}

The pump-probe response Eq. \eqref{eq:XiPP_definition} can be considered as the difference in linear responses of the ``pumped'' state $e^{i \mu M^\alpha} \ket{0}$ and of the ground state. The key distinction is the presence of excited QPs in the pumped state that are not present in the unperturbed ground state. The leading-order processes to consider are those where the pump and probe pulses each only excite pairs of anyons at $t=0$ and $t=t_1$, which must then be annihilated at $t=t_2$. Ref. \cite{mcginley_pump_probe_prl} insightfully points out the importance of statistical phases in the long-time regime of the real-time signal of pump-probe spectroscopy, with certain assumptions in the semi-classical analytical treatment to make the problem more tractable, but a precise quantitative understanding of pump-probe responses in realistic systems remains incomplete. Motivated by this, we provide a comprehensive framework based on various numerical and analytical methods to elucidate the role of braiding statistics in real-time pump-probe responses, and we find a clear signature of braiding statistics especially in the early-to-intermediate time regime.

\begin{figure*}
    \centering
    \includegraphics[width=0.95\linewidth]{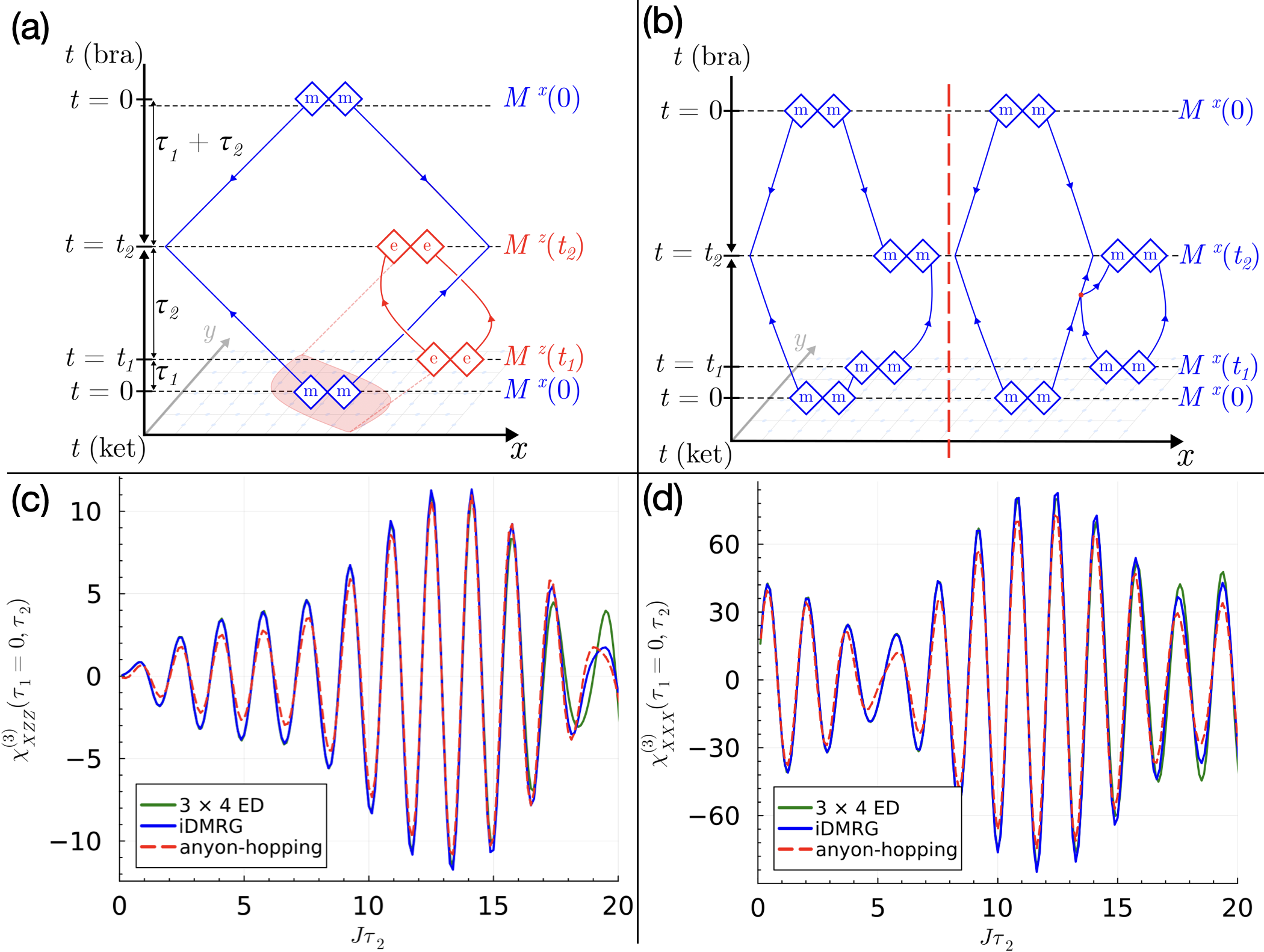}
    \caption{\textbf{$\chi^{(3)}$ Nonlinear Response:}
    Results here computed with $3\times 4$ ED and iDMRG with $J=1,h=0.05$, and with the anyon-hopping model on a $L_y$-periodic cylinder with $L_x=12, L_y=3$ with $\Delta_e=\Delta_m=2, h_e=h_m=0.05$; statistical interactions are accounted for by incorporating anyonic strings as in the right of Fig. \ref{fig:tc-lattice-diagram}.
    \textbf{(a)} Diagram of anyon braiding processes contributing to $\chi_{XZZ}^{(3)}$. A pair of $m$'s is created at $t=0$, followed by a pair of $e$'s created at $t=t_1$ and annihilated at $t=t_2$. The $e$'s must form a closed worldline loop, and the probability of braiding scales with the area of the region highlighted in red. 
    \textbf{(b)} Diagrams showing anyon processes which contribute to $\chi_{XXX}^{(3)}$. On the left, one $m$ created at $t=0$ is annihilated and replaced by a new $m$ at $t=t_1$, forming one large worldline loop. On the right, operators at $t=0$ and $t=t_1$ create distinct pairs of $m$'s, which can scatter off of each other before annihilating in pairs. 
    \textbf{(c)} Plot of $\chi_{XZZ}^{(3)}$ computed with ED, iDMRG and the anyon-hopping model. The linear-in-$\tau_2$ growth at early time results from the growing probability of braiding between $e$'s and $m$'s excited by the pump and probe pulses. 
    \textbf{(d)} Plot of $\chi_{XXX}^{(3)}$ computed with ED, iDMRG and the anyon-hopping model. The early time signal decays, similar to the linear response, due to the decreasing probability of forming closed QP worldline loops. }
    \label{fig:chi3-plots}
\end{figure*}

\bigskip

\noindent {\bf {XZZ Response}}: We first consider the case where $\alpha=X, \beta=\gamma=Z$ and examine the QP processes contributing to the response. Both terms in Eq. \eqref{eq:chi2_definition} include the operator $M^x(0)$ which excites two $m$'s but no operator to annihilate them, so we expect $\chi^{(2)}_{XZZ}$ to play little part in the observed $\Xi^{PP}_{XZZ}$ and focus our studies on $\chi^{(3)}_{XZZ}$. The second and third terms of Eq. \eqref{eq:chi3_definition} partly suffer from the same problem; this leaves only the first term of Eq. \eqref{eq:chi3_definition} to contribute, wherein $M^x(0)$ excites two $m$'s and $M^z(t_1)$ excites two $e$'s which annihilate by $M^z(t_2)$. The $e$'s must form a closed worldline loop, which allows for the probability that previously excited $m$'s can braid with this loop. In the pump-probe function $\Xi^{PP}_{XZZ}$, processes wherein the $m$'s do not braid are simply directly proportional to the linear response and thus are removed by the subtraction in Eq. \eqref{eq:XiPP_definition}; however, when an $m$ does intersect the worldline loop of the $e$'s, as shown in Fig. \ref{fig:chi3-plots}(a), the braiding leads to a relative phase shift between the two terms in Eq. \eqref{eq:XiPP_definition}, allowing these processes contribute to $\chi^{(3)}_{XZZ}$. The amplitude of $\chi^{(3)}_{XZZ}$ is proportional to the probability of braiding occurring, which in turn scales with the area enclosed by the $e$ worldlines, and thus we expect the amplitude of $\chi^{(3)}_{XZZ}$, and thus the amplitude of $\Xi^{PP}_{XZZ}$, to be an increasing function of $\tau_2$.

\bigskip

\noindent {\bf {Comparison of XZZ and XXX Responses}}: Next, we compare this behavior to the case where $\alpha=\beta=\gamma=X$. $\chi^{(2)}_{XXX}$ contains processes which create a large connected worldline loop which was not allowed in the $XZZ$ case, shown in Fig. \ref{fig:chi2-plots}(b), and thus $\chi^{(2)}_{XXX}$ has an appreciable contribution to the pump-probe response. $\chi^{(3)}_{XXX}$ includes many of the same types of processes as $\chi^{(3)}_{XZZ}$, simply replacing $e$'s with $m$'s, but whether a pump QP braids with the probe QPs is irrelevant (since all particles are mutually bosonic), and these terms are removed by the subtraction in Eq. \eqref{eq:XiPP_definition}, leaving no signal. Instead, additional processes like those shown in Fig. \ref{fig:chi3-plots}(b) contribute to $\Xi^{PP}_{XXX}$; these include processes like those on the left, where $M^x(t_1)$ and $M^x(t_2)$ annihilate and recreate single $m$'s to form one large worldline loop, and those on the right, wherein pump and probe QPs complete separate loops, but there is a chance for one QP to scatter off another. These processes insist on forming closed worldline loops, so we expect the amplitude of the signal to mimic the decaying behavior of the linear response.  The distinction between the amplitude in the braiding case $\chi^{(3)}_{XZZ}$ growing with time vs. the amplitude in the topologically trivial case $\chi^{(3)}_{XXX}$ decaying in time forms the key foundation of our result.

Figs. \ref{fig:chi3-plots}(c) and (d) show the behavior of $\chi^{(3)}_{\alpha,\beta,\gamma}(t)\equiv \chi^{(3)}_{\alpha,\beta,\gamma}(t_1=0,t_2)$ for both combinations of fields $\alpha,\beta,\gamma=X,Z,Z$ and $X,X,X$, each computed with ED and iDMRG. Again, we supplement this data with a model of hard-core anyons hopping on a lattice, and all three calculations agree quite well, with some deviation by the ED for $Jt\gtrsim15$ due to finite size effect in $L_x$ as estimated from Lieb-Robinson bound. In the $XZZ$ case, we clearly see the growth of the the signal amplitude at early times, as expected due to the braiding of pump and probe anyons; our further analysis of the anyon-hopping model elucidates that this is a linear-in-$t$ enhancement of the linear response $\chi^{(1)}_{ZZ}(t)$. In contrast, the $XXX$ case shows an initially decaying signal, similar to the linear response, due to the decreasing probability to form closed worldline loops, and our anyon-hopping analysis concretely demonstrates this similarity. We emphasize that the distinct braiding behavior is most apparent at early times, as the data become qualitatively similar for $Jt\gtrsim8$ due to finite size effect in $L_y$. 

\bigskip

\noindent {\bf {Analytical Understanding of $\chi^{(3)}$:}} 

Here we show the early time behavior of $\chi^{(3)}_{XZZ}$ receives a linear-in-$t$ enhancement relative to the linear response. In the early time regime (defined in detail below), we can approximate $m$'s created by $M^x(0)$ as static, since they have not drifted far from their initial positions. Consequently, the pump pulse creates a uniform background of $m$ particle pairs on different links; when hopping $e$ particles cross these links, the wave function changes sign due to the semionic statistics between $e$ and $m$ particles. Since $\Xi_{XZZ}^{PP}$ is defined as the difference between the hopping of $e$ particles with and without the $m$ particle background, we can establish a direct relation between the linear and pump-probe responses. Without the $m$ background, a world loop of $e$ particles composed of $d$ nearest-neighbor steps contributes an amplitude $h_e^{d}$ to the linear response. In the presence of $m$'s, those $e$ world loops that contribute to the pump-probe response must cross an $m$ link exactly once (otherwise they cancel after subtracting the linear response), replacing the $h_e$ by $-h_e$ at that link. For each $e$ world loop contributing weight $h_e^d$ to the linear response, the probability of crossing an $m$ link scales proportionally with the length of the loop $d$, and we therefore have a contribution to the pump-probe response $dh_e^{d-1}\times \delta h_e=2dh_e^d$ due to the subtraction in Eq. \eqref{eq:XiPP_definition}; however, this is conveniently achieved by the linear operator $2h_e\frac{d}{dh_e}$. Summing over these world loops yields the following relation for the pump-probe response in terms of the linear response:   
\begin{equation}
    \chi_{XZZ}^{(3)}(h_e,t)\approx 2h_e\frac{d}{dh_e}\chi_{ZZ}^{(1)}(h_e,t).
    \label{eq:xzz_relation}
\end{equation}
Using the simplified expression for the free boson linear response Eq. \eqref{eq:free-boson-chi1} in Eq. \eqref{eq:xzz_relation} captures the essential early time behavior and leads to
\begin{align}
    \chi^{(3)}_{XZZ,\text{Bessel}}(t)&\approx 4h_e\sin(2\Delta t)\frac{d}{dh_e}J_0^2(4h_et) \nonumber \\ 
    &=16h_et \sin(2\Delta t)\left(\frac{d}{dx}J_0^2(x)\right)|_{x=4ht},
    \label{eq:XZZ_Bessel}
\end{align}
where the last expression explicitly shows the linear-in-$t$ enhancement.

\begin{figure}
    \centering
    \includegraphics[width=0.95\linewidth]{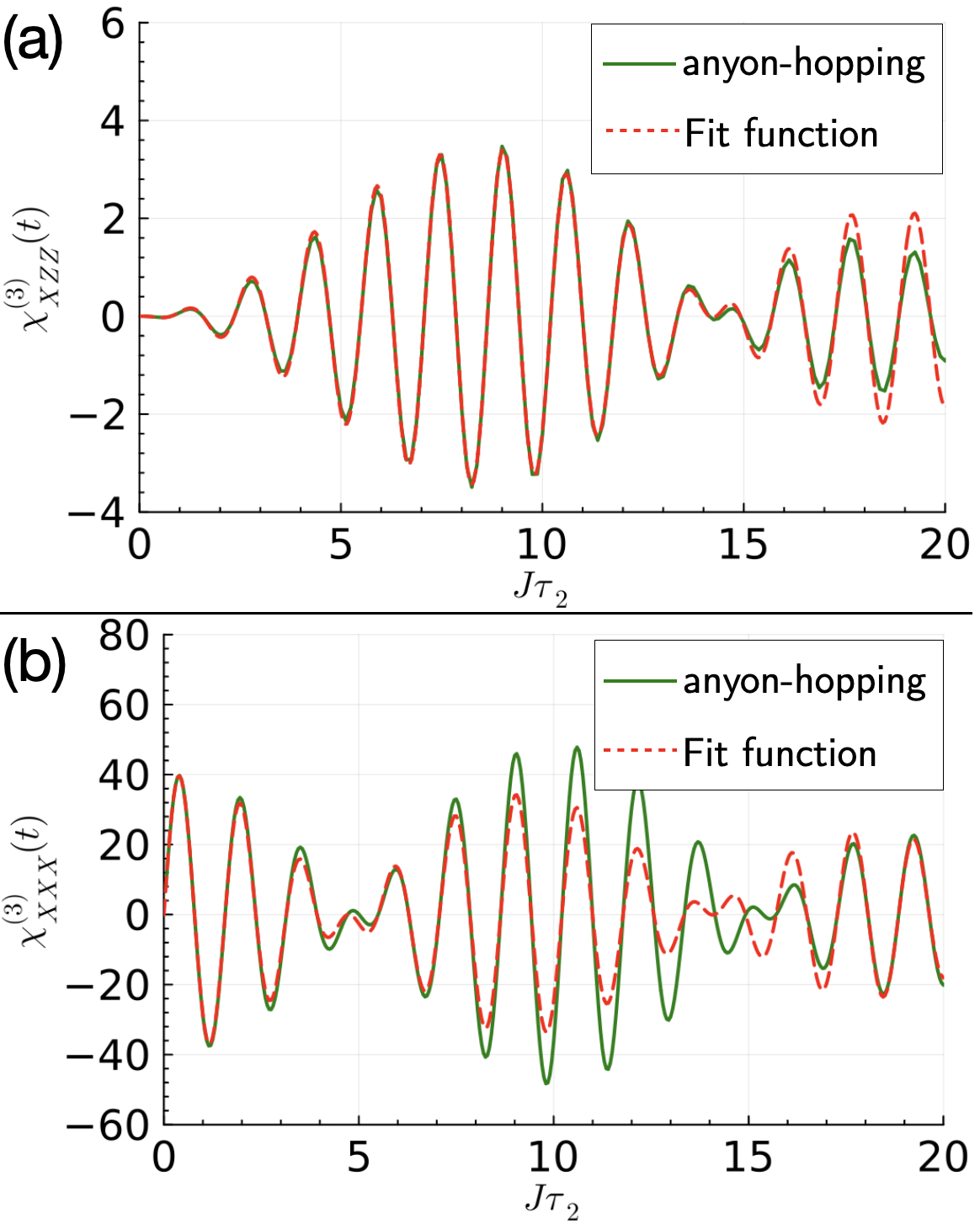}
    \caption{\textbf{$\chi^{(3)}$ Analytic Forms:}
    \textbf{(a)} Plot of $\chi^{(3)}_{XZZ}$ based on the anyon-hopping model and the analytical Bessel function form Eq. \eqref{eq:XZZ_Bessel}. The anyon-hopping data uses $h_e=h_m=0.05$ and $\Delta_e=\Delta_m=2$, while the Bessel form uses renormalized parameters $h_m=0.042$ and $\Delta_m=2$. The anyon-hopping model is done on a $15\times 15$ cylinder to ensure convergence.
    \textbf{(b)} Plot of $\chi^{(3)}_{XXX}$ based on the anyon-hopping model and the analytical Bessel function form Eq. \eqref{eq:XXX_Bessel}. In both cases we have used the bare parameters $h_m=0.05$ and $\Delta_m=2$. The anyon-hopping model is done on a $15\times 15$ periodic lattice to ensure convergence.}
    \label{fig:chi3-fitting}
\end{figure}

This result agrees very well at early times with the anyon-hopping simulation, as shown in Fig.\ref{fig:chi3-fitting}(a) albeit with a renormalized value of $h_e=0.042$. Although we have made several simplified assumptions, the universal features depend only on the kinetics of anyons and hold quite generally. We estimate the Bessel function description remains valid up until a time $Jt_{int}\sim j_{0,1}/(4h) \approx 14$, where $j_{n,k}$ denotes the $k^\text{th}$ zero of $J_n(x)$; this roughly measures the time for an $e$ particle to traverse one lattice constant, justifying our treatment of the $m$ particles as static. Effects of hard-core constraints and the statistical interaction between $e$ and $m$ particles also set in beyond this time scale. Microscopic details may change the prefactor in $t_{int}$, but the overall scaling relation $1/h_e$ remains generally valid. The long-time regime consists of several cycles of $t_{int}$, during which the hard-core constraints as well as statistical and non-statistical interactions between $e$ and $m$ particles become important. The linear-in-$t$ enhancement relation Eq. \eqref{eq:xzz_relation} assumes that $m$ particles are static, but in late times when $m$ particles are mobile, the phase space where $e$ particles can hop across $m$ strings is enhanced, resulting in a different power law enhancement, whose detailed study shall be left for future work.

\begin{widetext}
    The early-time behavior of $\chi^{(3)}_{XXX}$ is also well described by an analytical expression derived from the anyon-hopping model Eq. \eqref{eq:anyon_hopping_ham}. The magnetization operator $\sigma^x$ on a link between site $i$ and site $j$ can be written as $\sigma_{x}=v_{i}v_{j}+v_i^{\dagger}v_{j}+h.c.$. To illustrate the key structure of the full expression Eq. \eqref{eq:chi3_definition}, we compute a representative term, as all other terms yield the same result:$
        -i\sum_{{\delta}_{1,2,3,4}=\hat{x},\hat{y}}\ \sum_{j,l,m} \bra{0}[[[e^{iHt}v_0v_{0+\delta_1}e^{-iHt},v^{\dagger}_jv^{\dagger}_{j+\delta_2}],v_lv_{l+\delta_3}],v^{\dagger}_mv^{\dagger}_{m+\delta_4}]\ket{0}.$
\end{widetext}
Applying a Fourier transform and utilizing the relation $e^{iHt}v_{\textbf{k}}e^{-iHt}=e^{-i\epsilon_{\textbf{k}}t}v_{\textbf{k}}$, we obtain the following expression:
\begin{equation}
     \chi^{(3)}_{XXX}=\sum_{\textbf{k}}\text{sin}(2\epsilon_{\textbf{k}}t)(\text{cos}(k_x)+\text{cos}(k_y))^4.
\end{equation}
where the sum over four $\delta_j$'s leads to the factor $(\text{cos}(k_x)+\text{cos}(k_y))^4$, which can be rewritten as coming from four derivatives with respect to $x\equiv4h_mt$. With this in mind, summing over all different channels (corresponding to different creation and annihilation paths of $m$ particles) yields the full expression: 
\begin{equation}
    \chi^{(3)}_{XXX,\text{Bessel}}=\frac{160}{9}\text{sin}(2\Delta_m t)\frac{d^4J_0^2(x)}{dx^4}|_{x=4h_mt}.
    \label{eq:XXX_Bessel}
\end{equation}
The detailed functional form (omitted) consists of Bessel functions of orders $0,\ldots,4$. A combinatorial prefactor is verified by setting $h_m=0$ and comparing with the exact result $\chi^{(3)}_{XXX,h_m=0}=40\text{sin}(2\Delta_m t)$. 

Comparing Eq. \eqref{eq:XXX_Bessel} with numerical simulations of the anyon-hopping model as shown in Fig. \ref{fig:chi3-fitting}(b), we find excellent agreement up to the timescale set by the first zero of the Bessel function derivative $Jt\sim 1/(4h_m)=5$, beyond which deviations arise due to the hard-core constraints. One notable deviation is that from the asymptotic form of Bessel functions, Eq. \eqref{eq:XXX_Bessel} exhibits $1/t$ long-time decay. However, when hard-core constraints are incorporated, the decay is modified due to scattering between pump particles and probe particles and hence should be slower than $1/t$, highlighting the significant differences between the constrained and unconstrained cases. 

Ref. \cite{nandkishore2021spectroscopic} provides an analytical analysis of the $\chi^{(3)}_{XXX}$ for a toric code model in a field. However, their approach imposes hard-core constraints only at the level of the magnetization operator, while anyons are still allowed to evolve under a free Hamiltonian without such constraints. As we have pointed out above, ignoring hard-core constraints significantly affects the late-time behavior of pump-probe responses. Numerical and analytical calculations of the nonlinear responses based on mean-field theory of Kitaev spin liquids on the honeycomb lattice \cite{choi_theory_2020,nandkishore2021spectroscopic,qiang2024probing,kaib2025nonlinear} assume static gauge fluxes and therefore fail to correctly capture the winding phase between the Majorana fermions and dynamical gauge fluxes.

\bigskip

\begin{figure}
    \centering
    \includegraphics[width=0.95\linewidth]{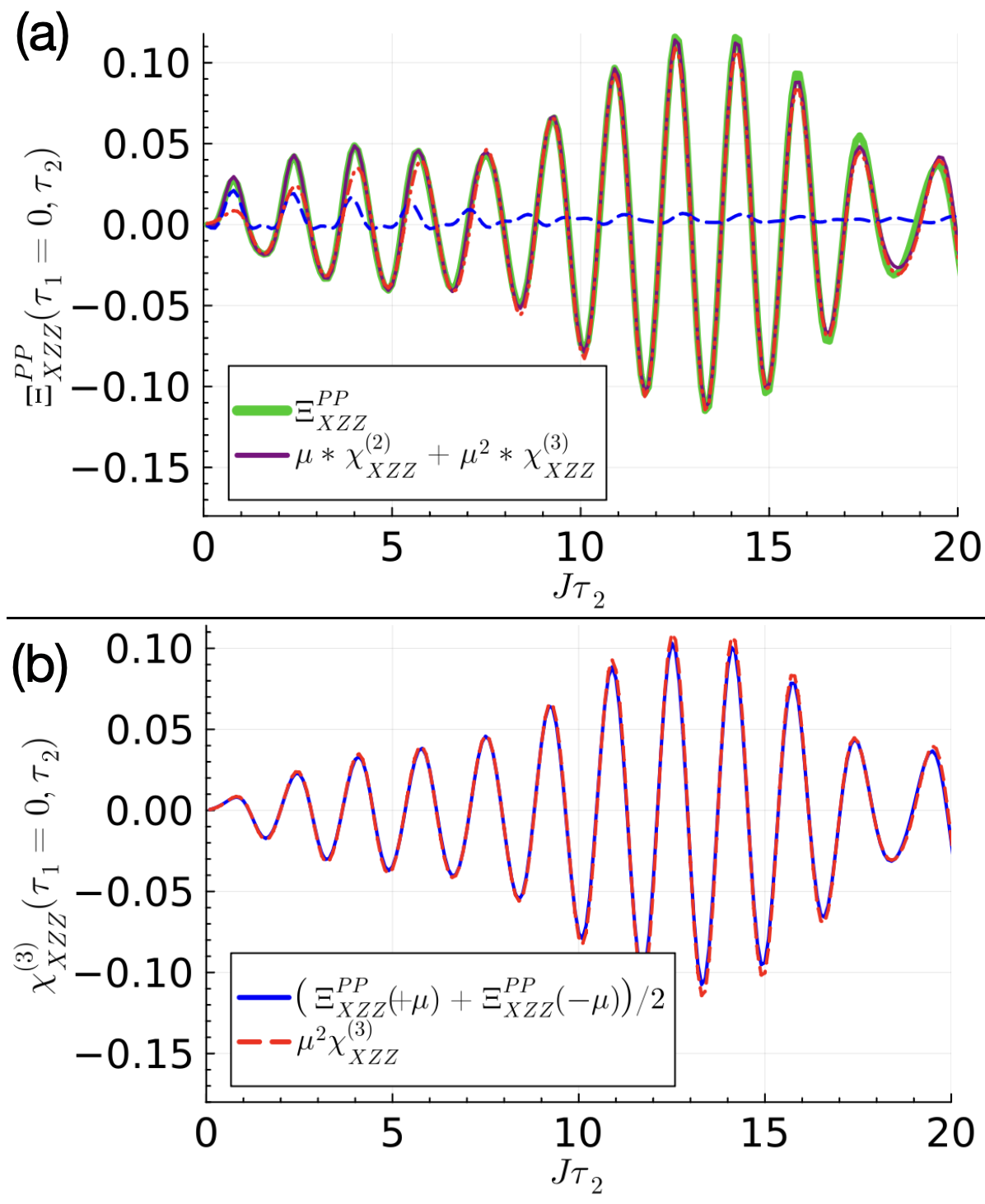}
    \caption{\textbf{Full Pump-Probe Response $\Xi^{PP}$:}
    \textbf{(a)} Plot of the full pump-probe response $\Xi_{XZZ}^{PP}$ with $J=1, h=0.05, \tau_1=0,$ and $\mu=0.1$. The blue and red dashed curves show $\mu\chi^{(2)}_{XZZ}$ and $\mu^2\chi^{(3)}_{XZZ}$ individually; their sum gives the purple curve $\mu\chi_{XZZ}^{(2)} + \mu^2\chi_{XZZ}^{(3)}$, which matches well with the exact calculation of $\Xi_{XZZ}^{PP}$ in bright green. Here we note that $\chi_{XZZ}^{(2)}$ is a small early time correction, and for $\tau_2\gtrsim 5$, $\chi_{XZZ}^{(3)}$ accurately captures most all of $\Xi_{XZZ}^{PP}$.
    \textbf{(b)} Plot of the combined full pump-probe response $\left(\Xi_{XZZ}^{PP}(+\mu) + \Xi^{PP}_{XZZ}(-\mu)\right)/2$ with $J=1, h=0.05, \tau_1=0$, along with $\mu^2\chi^{(3)}_{XZZ}$. Combining signals with opposite signs of $\mu$ eliminates the contributions of $\chi^{(2)}_{XZZ}$, making the braiding signature more apparent.
    }
    \label{fig:full_XiPP-plots}
\end{figure}

\noindent {\bf {Full pump-probe response:}}
To understand more precisely how these signals would appear in an experiment, we also exactly compute $\Xi^{PP}_{\alpha,\beta,\gamma}$ as defined in Eq. \eqref{eq:XiPP_definition} with ED using the same matrix exponential techniques as employed for the time evolution $e^{-i\mathcal{H}t}$ in order to compute the state $e^{i\mu M^\alpha}\ket{0}$ \cite{Haegeman_KrylovKit_2024}; these results are shown in Fig. \ref{fig:full_XiPP-plots}. We have plotted $\mu\chi^{(2)}_{XZZ}$, $\mu^2\chi^{(3)}_{XZZ}$, and $\Xi^{PP}_{XZZ}$ as computed with both Eqs. \eqref{eq:XiPP_definition} and \eqref{eq:Xi_expanded}. We see that $\chi^{(2)}_{XZZ}$ only contributes significantly to the very early peaks in $\Xi^{PP}_{XZZ}$. For $J\tau_2\gtrsim4$, $\chi^{(3)}_{XZZ}$ well describes the full behavior of $\Xi^{PP}_{XZZ}$. In the $XXX$ case, both $\chi^{(2)}_{XXX}$ and $\chi^{(3)}_{XXX}$ fail to separately capture the features of $\Xi^{PP}_{XXX}$, and their combined sum still features some variation from the exact result, suggesting contributions from some higher order terms. 

Our key result focuses on the distinction between behaviors of the third-order responses; by making the substitution $\mu\rightarrow-\mu$ in Eq.\eqref{eq:XiPP_definition} (physically, reversing the direction of the pump pulse polarization) and combining the signals with opposite signs, one can remove contributions from the odd powers of $\mu$, leaving only $\chi^{(3)}$ as the dominant signal, even in systems which allow a nonzero $\chi^{(2)}$. This is shown for $\chi^{(3)}_{XZZ}$ in Fig. \ref{fig:full_XiPP-plots}(b) in comparison to the combination $\Xi^{PP}_{XZZ}(+\mu)+\Xi^{PP}_{XZZ}(-\mu)$, where the contributions from $\chi^{(2)}_{XZZ}$ are completely removed, leaving behind solely $\chi^{(3)}_{XZZ}$ with excellent agreement. More detailed discussion and results for $\chi^{(2)}$ can be found in Appendix \ref{appendix:chi2}.

\bigskip

\noindent {{\bf Polarized phase responses:}}
\begin{figure*}
    \centering
    \includegraphics[width=0.95\linewidth]{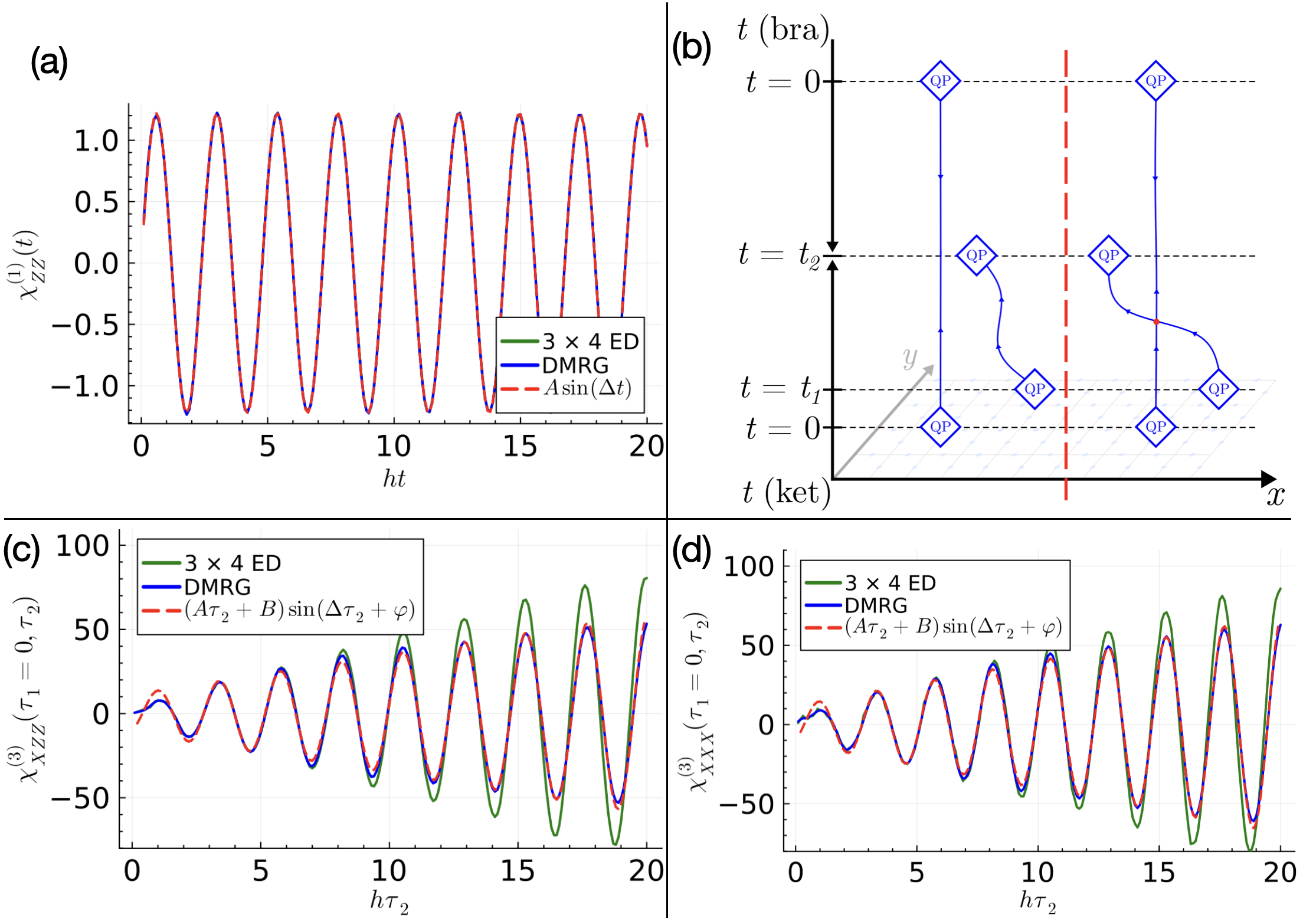}
    \caption{\textbf{Polarized Toric Code Results:}
    Results here computed with $3\times 4$ ED and finite-size DMRG with $L_x=19$ and $L_y=3$ for $J=0.2, h=1$.
    \textbf{(a)} Linear response $\chi^{(1)}_{ZZ}$, fitting the DMRG to a sine function $A\sin(\Delta t)$ with $A\approx1.2, \Delta\approx2.6$.
    \textbf{(b)} Diagrams showing the relevant QP processes contributing to $\chi^{(3)}$ in a polarized state. The QP created by the (weak) pump follows a semi-classical trajectory, while the QPs created at later times follow fully quantum trajectories, allowing them to have some probability to scatter off the pump QP. The scattering phase shift contributes to a nonzero pump-probe response which grows linearly in time (in proportion to the length of the QP worldlines).
    \textbf{(c)} Plot of $\chi_{XZZ}^{(3)}$, with the DMRG data fit to the form $(A\tau_2+B)\sin(\Delta \tau_2+\varphi)$ to capture both the static and dynamic QP responses, with $A\approx2.4, B\approx11.3, \Delta\approx2.6, \varphi\approx-1.0$. Note the ED data begins to diverge from the DMRG around $\tau_2\sim10$ due to finite size effects.
    \textbf{(d)} Plot of $\chi_{XZZ}^{(3)}$, with the DMRG data fit to the form $(A\tau_2+B)\sin(\Delta \tau_2+\varphi)$ with $A\approx2.8, B\approx11.8, \Delta\approx 2.6, \varphi\approx-0.9$; here we note that in the polarized phase, $\Xi^{PP}_{XZZ}$ and $\Xi^{PP}_{XXX}$ cannot be clearly distinguished, since they are described by the same QP dynamics.
    }
    \label{fig:polarized-TC-plots}
\end{figure*}
In order to fully understand the nonlinear responses of our model Eq. \eqref{eq:TC_Ham} and distinguish the braiding behavior of $\chi^{(3)}_{XZZ}$ in the toric code phase, we also present data in Fig. \ref{fig:polarized-TC-plots} of the linear and nonlinear responses in the field-polarized regime, choosing $h/J=5$. In this case, the operators $M^\alpha$ can create/annihilate single spin-flips which are hard-core bosons. In the linear response $\chi^{(1)}_{ZZ}$, $M^z(0)$ creates a QP that $M^z(t)$ annihilates, and we now just have an oscillation at the gap frequency due to the dynamical phase acquired. Unlike the anyonic case where the fractionalized nature of anyons leads to temporal decay in the linear response, local spin flips exhibit no such decay, indicating an absence of fractionalization. The second order responses vanish in the limit $h/J\rightarrow\infty$, due to an Ising-like symmetry along the $(\hat{x}+\hat{z})$ axis. Both third order responses, $\chi^{(3)}_{XZZ}$ and $\chi^{(3)}_{XXX}$, are qualitatively indistinguishable, since both arise due to QP processes like those shown in Fig. \ref{fig:polarized-TC-plots}(b). The process on the left is proportional to linear response and subtracted off, while the process on the right features a scattering between pump and probe QPs. The scattering phase shift acquired as a result can lead to a nonzero contribution to the pump-probe response, and we expect the signal to grow in time as the probability for scattering grows proportionally to the probe QP worldline length. This is essentially the two-dimensional generalization of the argument and conclusion of Ref. \cite{fava_divergent_2023}, where they found that QP scattering phase shifts in one-dimensional systems lead to a linear-in-time divergence of the pump-probe response.

The linear response data for the polarized fits well to a form $A\sin(\Delta t)$. Both nonlinear responses fit well to a form with a linear-in-time divergence, $(A\tau_2+B)\sin(\Delta \tau_2 + \varphi)$; the term proportional to $t$ accounts for the scattering phase shift, while the constant term accounts for processes where QPs are approximately static and improves the fit at early times. The fits for all the data shown agree in their value for the gap frequency $\Delta$. We emphasize that the $\chi^{(3)}$ nonlinear responses in the anyonic toric code and polarized toric code both feature linear-in-time enhancements at early times, so one needs to needs to simultaneously observe the decaying linear response (due to fractionalization) and pulse polarization distinctions (due to anyon selection rules) in order to support the linear-in-time enhancement as a signature of braiding.

\section{Time scales}
Based on our analytical treatment of the pump-probe response, we identify a key time scale central to our results: the intermediate time scale $t_{\text{int}}$, which we estimate as $t_{\text{int}} \equiv \pi / (4h_{\text{kinetic}})$, derived from the average spacing between the zeros of the Bessel function $J_0$. $\chi^{(3)}_{XZZ}$ and $\chi^{(3)}_{XXX}$ most clearly differ for $t<t_\text{int}$, with $\chi^{(3)}_{XXX}$ showing decay in time resembling linear response and $\chi^{(3)}_{XZZ}$ showing linear-in-$t$ enhancement relative to linear response. The hard-core constraints cause $\chi^{(3)}_{XXX}$ to significantly deviate from the free boson analysis also on a time scale $\propto t_\text{int}$. In our specific case $Jt_{\text{int}}\sim 15$, in accordance with our numerical observation. Using the estimated Lieb-Robinson velocity $v_{LR} \sim 2h_{\text{kinetic}}$, we obtain a model-independent system size $2v_{LR}t_{\text{int}}\approx 3$ where the distinct behavior of $\chi^{(3)}_{XZZ}\text{ versus }\chi^{(3)}_{XXX}$ becomes evident, which justifies our use of ED and iDMRG. As seen in the $\chi^{(3)}_{XZZ}$ data, finite-size effects set in around $Jt \sim 8$, indicating an underestimation of $v_{LR}$. For $Jt\lesssim8$ iDMRG and ED can unambiguously distinguish between anyons with and without braiding in $\chi^{(3)}_{XZZ}\text{ versus }\chi^{(3)}_{XXX}$.

This also imposes stringent constraints on system size. For example, to clearly observe five cycles of slow modes described by the Bessel functions, a system with linear size at least $5\times (2v_{LR} t_{\text{int}})\sim 16$ is required, well beyond the capability of most existing numerical techniques and might only be achievable within the anyon-hopping model. At long times order of magnitude larger than $t_\text{int}$, pump anyons drift sufficiently far apart such that the static anyon approximation no longer holds. In this regime, the phase space for braiding can be enhanced, leading to a different scaling than we study here \cite{mcginley_pump_probe_prl,mcginley2024anomalous}.  While our anyon-hopping model only strictly describes the case where anyons have no interactions other than their mutual statistics and hard-core constraints, these models are expected to exhibit the correct long-time behavior, where responses mainly depend on far-apart anyons and nonlocal physics. We leave the precise numerical study of the long-time regime for future work.

\section{Methods}
\label{section:methods}

\subsection{Exact diagonalization} 
To efficiently compute exact energies and eigenstates of the $L_x\times L_y$ toric code using exact diagonalization, we exploit translation invariance in both directions to construct a basis of states that are symmetry eigenstates, then build the Hamiltonian and magnetization operators $M^{\alpha}$ from their matrix elements in this basis \cite{sandvik_2010}. Since both the Hamiltonian and $M^{\alpha}$ respect all symmetries, it suffices to work within the zero-momentum sector of the Hamiltonian, which contains the ground state. Time evolution can be efficiently performed using Krylov subspace projection methods to approximate the action of $\exp(-i \mathcal{H} t)$ on a state to a desired precision \cite{expokit1998,Haegeman_KrylovKit_2024}. These techniques allow us to efficiently simulate a toric code with $L_x=3, L_y=4$ (for a total of $2L_xL_y=24$ sites) and compute all the relevant correlation functions. We have also used ED results to access high-resolution signals in frequency space simply by numerically Fourier transforming the time-domain signals.

\subsection{MPS-based methods} 
We use infinite DMRG (iDMRG) developed using Julia's ITensors package \cite{ITensor,ITensor-r0.3} to obtain the $\mathbb{Z}_2$ spin liquid ground state of the toric code on a cylinder with $L_x \times L_y$ sites, where $L_x$ is infinite and $L_y = 3$ is periodic. iDMRG optimizes only a single unit cell of size $L_y$ in each sweep, making it more efficient than finite DMRG, and translational symmetry along $x$ direction is preserved to reduce computational cost in time-evolution. After obtaining the infinite MPS ground state from iDMRG, we construct a finite chain with ``infinite" boundary conditions from many iterated unit cells along with two effective boundary sites to represent the semi-infinite chains \cite{phien2012infinite, milsted2013variational}. We retain enough unit cells along the $x$-direction to ensure that perturbations applied at the center do not effect the boundary sites within the simulation time. We then apply local quenches and evolve the system in time using single-site TDVP, with a time step $\delta t = 0.125$ and dynamic bond dimension growth via global subspace expansion. Exploiting the translational symmetry of the toric code ground state, we reduce the total magnetization $M^{\gamma}(t_2)$ to $\sigma^{\gamma}(t_2)$ within a single unit cell at the center of the cylinder. To mitigate rapid entanglement growth following global quenches in the $\mathbb{Z}_2$ spin liquid phase, we rewrite the response function Eq.~\eqref{eq:chi3_definition} as a sum of local four-point correlators, each computed separately via TDVP. Translational symmetry along $x$ direction is used wherever possible to further reduce computational cost. For polarized phases, where entanglement grows slowly, we instead use finite DMRG with finite $L_x$, perform a global quench, and check convergence with increasing $L_x$.

\subsection{Anyon-hopping model} 
We have also computed the pump-probe response using an anyon hopping model \cite{hatsugai1991braid,hatsugai1991anyons,kirchner2023numerical}, where hard-core anyons hop on an $L_x \times L_y$ lattice, as shown in Fig. \ref{fig:tc-lattice-diagram}. We restrict the dynamics to subspaces with fixed particle number, enforcing hard-core constraints that prevent anyons of the same type from occupying the same site. Exploiting ranslation symmetry further reduces the Hilbert space dimension. For $\chi^{(3)}_{XXX}$ response involving four $m$ anyons we work on a torus to exploit translation symmetry along both directions. For $\chi^{(3)}_{XZZ}$ response which involves two $e$ anyons and two $m$ anyons, we work on a cylinder to utilize translation symmetry along $L_y$ direction. Time evolution is carried out using the $exponentiate$ functionality provided by Julia's KrylovKit package \cite{Haegeman_KrylovKit_2024}.

\section{Summary and outlook}

We show that nonlinear pump-probe spectroscopy on a quantum spin liquid model performed with magnetic fields in different directions allows us to detect the braiding of fractionalized anyons. Contrasting a control setup with the pump, probe, and measurement magnetic fields in the same direction (XXX) against a setup with magnetic fields that involve non-commuting operators (XZZ) highlights the non-trivial signature of anyon braiding in the early time dependence; $\chi^{(3)}_{XZZ}$ grows with time in contrast to a time decay for the control case $\chi^{(3)}_{XXX}$. We have so far demonstrated the braiding signatures for the toric code, but the methods and analysis can be generalized to other frustrated spin models pertinent to quantum spin liquids. 

We reemphasize the analysis should be performed in real time, as opposed to 2D coherent spectroscopy which Fourier transforms the time-domain data, since the signal in frequency space obscures the distinctive features of the XZZ and XXX nonlinear response at early to intermediate times. We propose the following general outline for an experimental study that can view our proposed braiding signature:
\begin{enumerate}
    \item Measure the linear response Eq.\eqref{eq:lin_definition}, looking for a decaying time-domain signal and broad frequency-domain signal as evidence of fractionalization.
    \item If fractionalization is observed, measure the pump-probe response Eq.\eqref{eq:XiPP_definition}, then remeasure with reversed pump polarization (taking $\mu\rightarrow-\mu$)
    \item Combine the pump-probe signals to remove signals odd in $\mu$ to look for the linear-in-time enhancement at early times as evidence of braiding.
\end{enumerate}
To address the challenge of finding ideal QSL materials, experiments should focus on characterizing QSL candidates with a considerably suppressed magnetic ordering temperature $T_c$ from the Curie-Weiss temperature $\theta_{CW}$. Nonlinear response in the window $\theta_{CW} < T < T_c$ is dominated by quantum fluctuations and should show strong deviations from linear response in that temperature range.

\acknowledgments
XY acknowledges the hospitality of Prof. Gang Chen at Peking University, where part of the work was completed. This research was partially supported by the Center for Emergent Materials, an NSF MRSEC, under award number DMR-2011876 (XY) and from NSF-DMR Grant No. 2138905 (RB, NT).

\appendix
\section{Overview of Nonlinear Response Formalism}
\label{appendix:nonlinear_review}

We consider a system with a time-dependent Hamiltonian $\mathcal{H}(t) = \mathcal{H}_0 + V(t)$ where the perturbation $V(t)$ is of the form $-\vec{B}(t)\cdot \vec{M}$. For now, we work in the interaction picture where the state of the system is described by the density operator $\rho_I(t)$, which evolves according to the Liouville equation
\begin{equation}
    \frac{\partial \rho_I}{\partial t} = i[\rho_I(t), V_I(t)]
\end{equation}
and operators evolve like $\mathcal{V}_I(t) = e^{i\mathcal{H}_0t} \mathcal{V} e^{-i\mathcal{H}_0t}$. 
    \begin{widetext}
    We assume the system is in its ground state at the time $t_0$ and expand the density operator as $\rho_I(t) = \sum_{n=0}^\infty \rho_I^{(n)}(t)$ where:
    \begin{align}
        \rho_I^{(0)}(t) &= \ket{0}\bra{0} \equiv \rho(t_0) \\
        \rho_I^{(1)}(t) &= i\int_{t_0}^{t} dt_1 [\rho(t_0), V_I(t_1)] \\
        \rho_I^{(n)}(t) &= i^n \int_{t_0}^t dt_n \int_{t_0}^{t_n} dt_{n-1} \cdots \int_{t_0}^{t_2} dt_1 [[[[\rho(t_0), V_I(t_1)], ...], V_I(t_{n-1})], V_I(t_n)].
    \end{align}
    The induced magnetization can also be expanded as $M^\alpha(t) = \sum_{n=0}^\infty M^{\alpha, (n)}(t)$ with each term in this sum given by 
    \begin{align}
        &M^{\alpha, (n)}(t) = \Tr\left(\rho^{(n)}_I(t), M^\alpha_I(t)\right) \nonumber \\
        &= (-i)^n \int_{t_0}^t dt_n \int_{t_0}^{t_n} dt_{n-1} \cdots \int_{t_0}^{t_2} dt_1 \Tr\big( [[[\rho(t_0), M^{\alpha_1}_I(t_1)], ...], M^{\alpha_n}_I(t_n)] M^{\alpha}_I(t) \big) \nonumber \\
        &\times B^{\alpha_1}(t_1)...B^{\alpha_n}(t_n) \nonumber \\ 
        &= (-i)^n \int_{t_0}^t dt_n \int_{t_0}^{t_n} dt_{n-1} \cdots \int_{t_0}^{t_2} dt_1 \Tr\big( \rho(t_0) [M^{\alpha_1}_I(t_1),[..., [ M^{\alpha_n}_I(t_n), M^{\alpha}_I(t)]]] \big) \nonumber \\
        &\times B^{\alpha_1}(t_1) ... B^{\alpha_n}(t_n) 
    \end{align}
    where we have substituted in the form of $V_I(t)$ mentioned previously and used the fact that $\Tr\big(A[B,C]\big) = \Tr\big([A,B]C\big)$. We can then write
    \begin{equation}
        M^{\alpha,(n)}(t) = N\int_{t_0}^t dt_n \int_{t_0}^{t_n} dt_{n-1} \cdots \int_{t_0}^{t_2} dt_1 \chi_{\alpha_1, ..., \alpha_n, \alpha}^{(n)}(t_1, ..., t_n,t) B^{\alpha_1}(t_1) ... B^{\alpha_n}(t_n)
    \end{equation}
    where we have defined the $n$\textsuperscript{th}-order susceptibility 
    \begin{align}
        \chi_{\alpha_1, ..., \alpha_n, \alpha}^{(n)}(t_1, ..., t_n,t) = \frac{(-i)^n}{N} &\Tr\big( \rho(t_0) [M^{\alpha_1}_I(t_1),[..., [ M^{\alpha_n}_I(t_n), M^{\alpha}_I(t)]]] \big) \nonumber \\ 
        &\times \Theta(t_1-t_0)...\Theta(t_n-t_{n-1}) \nonumber \\
        = \frac{(-i)^n}{N} &\langle 0 | [M^{\alpha_1}_I(t_1),[..., [ M^{\alpha_n}_I(t_n), M^{\alpha}_I(t)]]] |0 \rangle \nonumber \\
        &\Theta(t_1-t_0)...\Theta(t_n-t_{n-1}) 
    \end{align}
    with the step functions $\Theta(t)$ to enforce causality.
    \end{widetext}

The main text uses a slightly simplified notation to correspond with the specific experimental setup considered. The function defined in the text as $\chi^{(1)}_{\alpha,\beta}(t)$ is the same as $\chi^{(1)}_{\alpha,\beta}(0,t)$; the forms presented in this appendix for $\chi^{(n\geq 2)}$ can be reduced to the forms in Eqs. \eqref{eq:chi2_definition} and \eqref{eq:chi3_definition} by setting the first $n-1$ time variables to 0 and replacing $t_n\rightarrow t_1$, and $t \rightarrow t_2$. The results in the text are presented considering the case where the first two pulses are concurrent ($\tau_1=t_1=0$), in which case the time delay between the second pulse and measurement $\tau_2=t_2$ becomes the only time variable.

\section{2\textsuperscript{nd}-order Response}
\label{appendix:chi2}

\begin{figure*}
    \centering
    \includegraphics[width=0.95\linewidth]{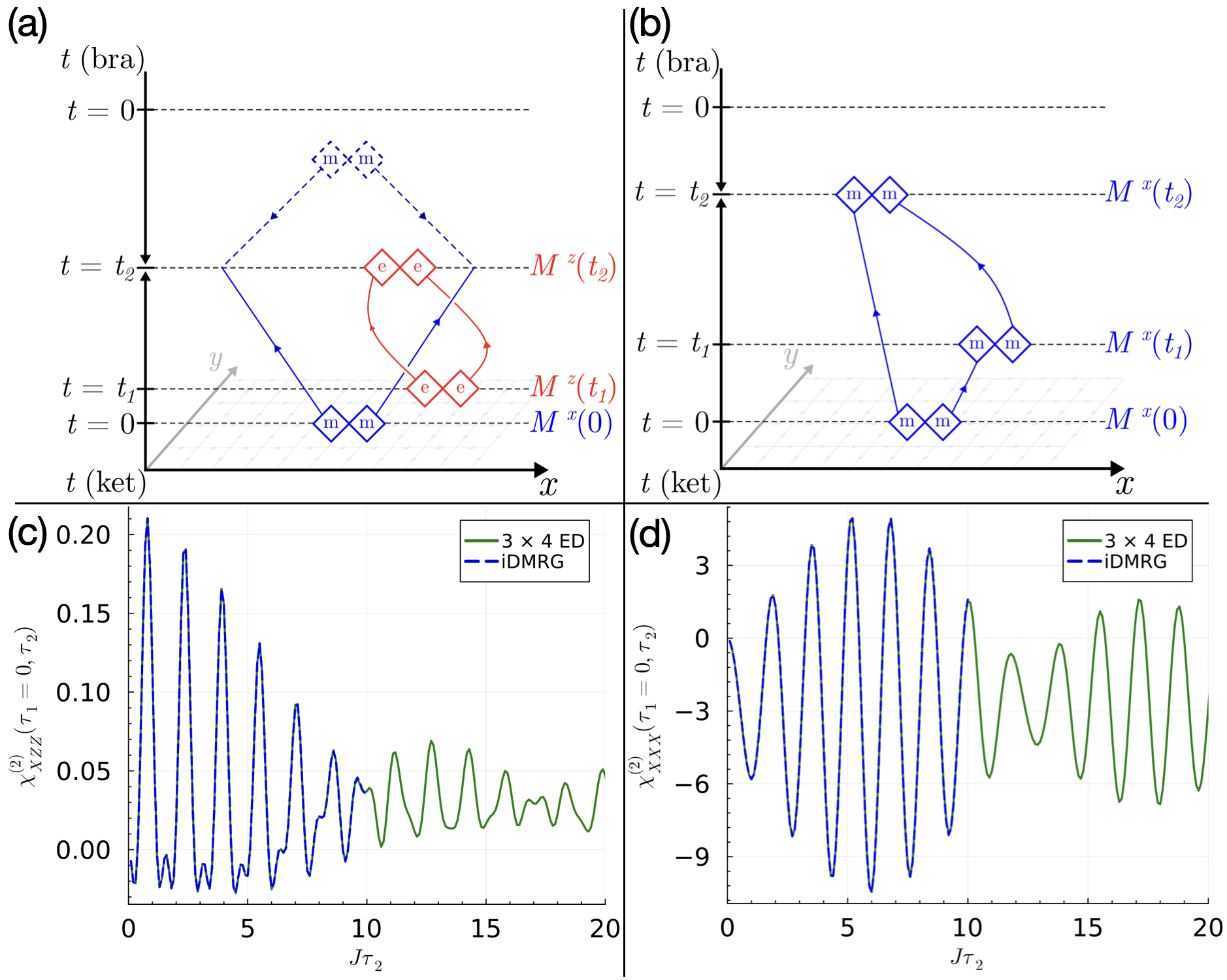}
    \caption{\textbf{$\chi^{(2)}$ Nonlinear Response:}
    \textbf{(a)} Diagram showing the relevant process contributing to $\chi_{XZZ}^{(2)}$. The $m$'s created at $t=0$ are only created on the ket side of the response, meaning that they must decay virtually since no other operator to annihilates them. As a result, we expect this signal completely vanishes in the thermodynamic limit since the $m$'s can spread arbitrarily far apart in an infinite size system.
    \textbf{(b)} Diagram showing the relevant process contributing to $\chi^{(2)}_{XXX}$. One of the $m$'s created at $t=0$ is annihilated and recreated by $M^x(t_1)$ before both $m$'s are annihilated together by $M^x(t_2)$, forming a large worldline loop that contributes to a decaying form for $\chi^{(2)}_{XXX}$. 
    \textbf{(c)} Plot of $\chi_{XZZ}^{(2)}$ for $J=1, h=0.05, \tau_1=0$ computed with ED and iDMRG. 
    \textbf{(d)} Plot of $\chi_{XXX}^{(2)}$ for $J=1, h=0.05, \tau_1=0$ computed with ED and iDMRG. In this case, there are allowed QP processes that annihilate all the $m$'s created, so this signal plays a significant role in the observed experimental response.
    }
    \label{fig:chi2-plots}
\end{figure*}

We show the results for $\chi^{(2)}_{\alpha,\beta,\gamma}(t_1=0,t_2)$ for both $XZZ$ and $XXX$ cases in Figs. \ref{fig:chi2-plots}(c) and (d). We see that the initial amplitude of $\chi^{(2)}_{XZZ}$ is small (relative to that of $\chi^{(2)}_{XXX}$) and quickly decays. This signal relies on the virtual annihilation of the $m$'s (shown in Fig. \ref{fig:chi2-plots}(a)) and therefore is largest at very short times, when the $m$'s have not spread far from each other and can more easily annihilate by chance. In the thermodynamic limit, we expect this signal may vanish more quickly than in our simulations, since the small finite size may be reinforcing the annihilation of $m$'s that would have otherwise traveled far apart. In contrast, $\chi^{(2)}_{XXX}$ has a much larger amplitude initially and an overall decaying profile due to the need of $m$'s to form closed worldline loops (shown in Fig. \ref{fig:chi2-plots}(b)) and the decreasing probability of doing so at larger $\tau_2$. 

\section{Polarized TFIM}
\label{appendix:tfim}

The transverse-field Ising model (TFIM) is a simple spin model consisting of nearest-neighbor Ising couplings along the $z$-direction with an applied magnetic field along the $x$-direction, as described by the Hamiltonian
\begin{equation}
    \mathcal{H}_{TFIM} = -J\left( \sum_{\langle j,k\rangle} \sigma_j^z \sigma_j^z + g \sum_j \sigma_j^x \right).
\end{equation}
We consider this model on a 2-dimensional square lattice in the paramagnetic phase where $g\gg 1$: the QP excitations of this phase are single spin-flips created by the action of $\sigma_z$ on a site, so we investigate the linear and nonlinear response of this system coupled to the magnetization $M^z$. 

\begin{figure}
    \centering
    \includegraphics[width=0.95\linewidth]{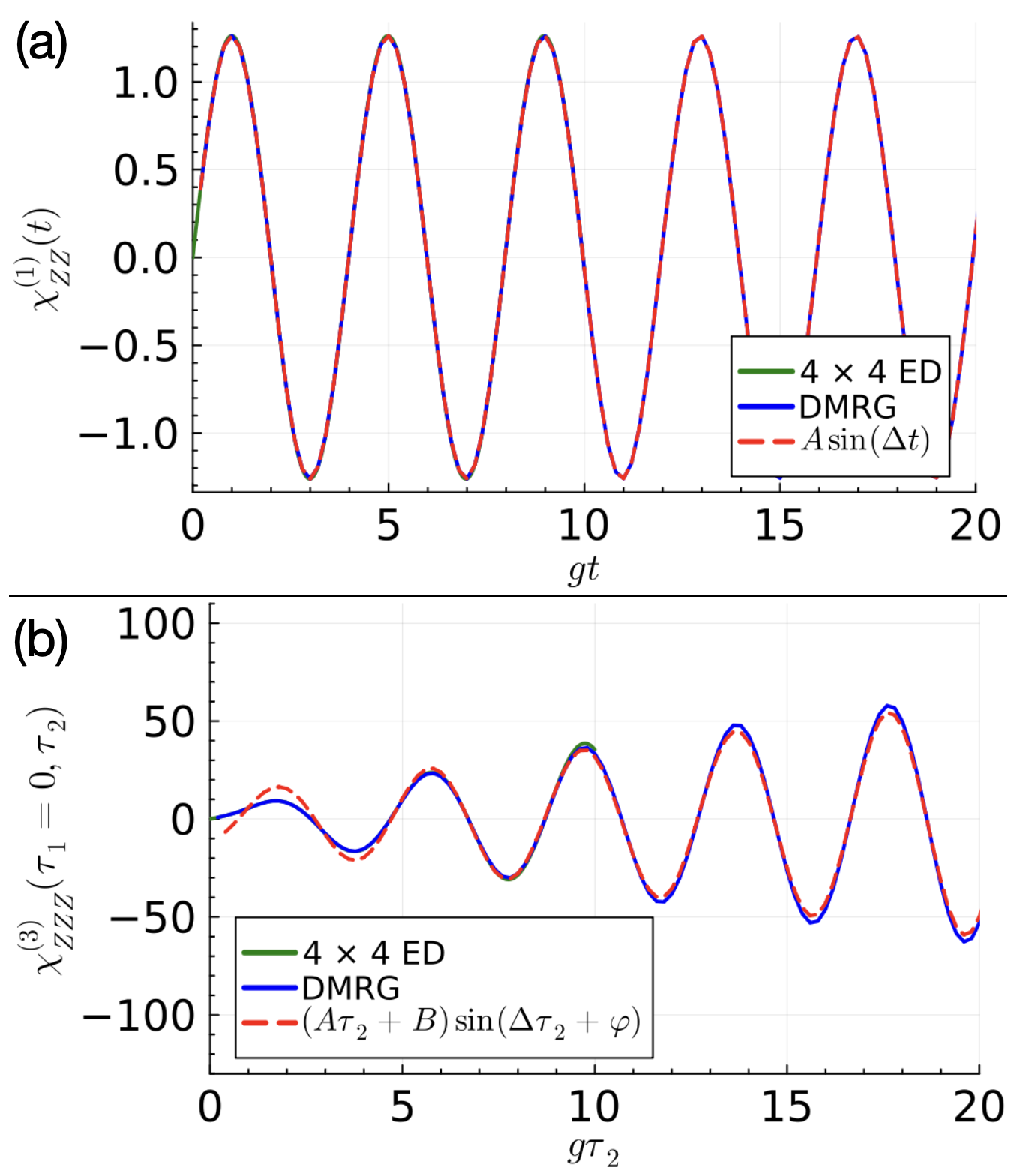}
    \caption{\textbf{Polarized TFIM Results: }Results here obtained using $4\times 4$ ED and finite DMRG with $L_x=29,L_y=4$ for the polarized TFIM with $J=0.1,g=1$.
    \textbf{(a)} Plot of the polarized TFIM linear response $\chi_{ZZ}^{(1)}$ computed with ED and DMRG, with DMRG data fit to $A\sin(\Delta t)$ with $A\approx1.3,\Delta\approx1.6$. This follows the expected sinusoidal behavior in agreement with the polarized toric code phase.
    \textbf{(b)} Plot of the polarized TFIM nonlinear response $\chi_{ZZZ}^{(3)}$ computed with ED and DMRG, with DMRG data fit to $(A\tau_2+B)\sin(\Delta\tau_2+\varphi)$ with $A\approx-2.4,B\approx-12.2,\Delta\approx1.6, \varphi\approx2.0$, which agrees well with the oscillating linear-in-time growth of the polarized toric code phase.
    }
    \label{fig:tfim-plots}
\end{figure}

The linear response $\chi^{(1)}_{ZZ}$, shown in Fig. \ref{fig:tfim-plots}(a), is simply a sinusoidal oscillation with frequency equal to the gap to create these spin-flip excitations. $\chi^{(2)}_{ZZZ}$ is zero for all times due to the Ising $z$-spin-flip symmetry. The behavior of $\chi^{(3)}_{ZZZ}$, shown in Fig. \ref{fig:tfim-plots}(b), is dominated by the same type of QP scattering as in the polarized toric code, with a linear-in-time divergence due to the increasing probability of braiding scaling with the length of QP worldlines. This can be understood as follows: in the limit where $g\gg 1$, the ground state is polarized along $x$ direction. The linear response is represented as $\chi^{(1)}_{zz}(t)\equiv \text{Im}\sum\limits_j\bra{0}[\sigma_{z,o}(t),\sigma_{z,j}]\ket{0}$. If we expand the operator commutator $\sum\limits_j[\sigma_{z,o}(t),\sigma_{z,j}]$ in powers of $t$, then only terms that are composed of $\sigma_x$ strings would contribute to $\chi^{(1)}$. Therefore formally we can do the following expansion:
\begin{equation}
    \sum\limits_j[\sigma_{z,o}(t),\sigma_{z,j}]\sim i\sum\limits_{n,l}\frac{t^n}{n!}D_{x,l}(n)+(...),
\end{equation}
where $D_{x,l}(n)$ is a string of $\sigma_x's$ with length $\mathcal{O}(n)$, and $l$ labels different configurations. The $(...)$ part contains operator strings with $\sigma_y$ or $\sigma_z$, so they have negligible contributions to the linear response. In general parts with one $\sigma_y$ and two $\sigma_y$s exist that contribute to $\chi^{(2)}$ and $\chi^{(3)}$, but only change the results quantitatively.

Consider now the pump-probe response $\chi^{(3)}(t)\equiv \text{Im}(\bra{0}[[[\sigma_{z,0}(t),\sigma_z],\sigma_z],\sigma_z]\ket{0})$. Focusing on terms originated from $D_{x,l}(n)$, we find that the two out-most commutators with $\sigma_z$ must be positioned at the same position to get a string that is still only composed of $\sigma_x$'s; therefore, 
\begin{equation}
    [[[\sigma_{z,0}(t),\sigma_z],\sigma_z],\sigma_z]\sim i\sum\limits_{n,l}n\frac{t^n}{n!}D_{x,l}(n)+(...),
\end{equation}
where $(...)$ includes negligible contributions. We now have $\chi^{(3)}(t)\sim t\frac{d}{dt}\chi^{(1)}(t)$. Since $\chi^{(1)}(t)$ is of the form $\text{sin}(2\Delta t)$, we expect $\chi^{(3)}(t)$ to be of the form $t\text{cos}(2\Delta t)$, which matches very well with the fitting. This argument places the physical picture of pump-particle scattering with a probe particle, as shown in Fig. \ref{fig:polarized-TC-plots}(b), on a more quantitative foundation.

\section{Fourier and Spectral Analysis}
\label{appendix:2d-spectra}

\begin{figure}
    \centering
    \includegraphics[width=0.95\linewidth]{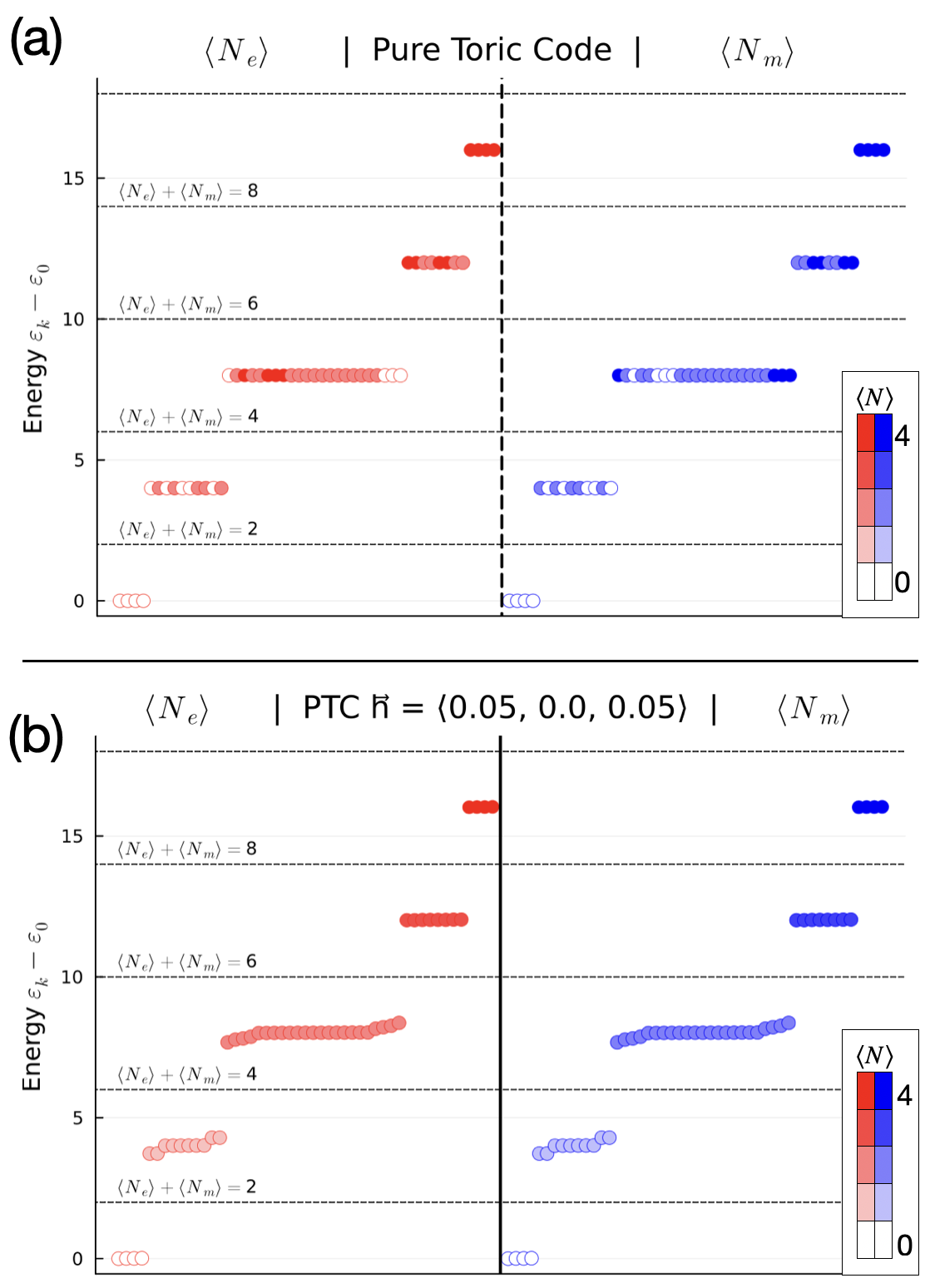}
    \caption{\textbf{Toric Code Eigenspectra:}
    \textbf{(a)} Spectrum of the pure toric code model with $J=1, h=0$ computed with ED on a $2\times 2$ system for clarity. Each half of the plot both show the same spectrum, with the red saturation on the left half indicating the number of $e$ QPs in the state, and the blue saturation on the right indicating the number of $m$ QPs. The spectrum is clearly highly degenerate and sorted into sectors with the same total number of QPs.
    \textbf{(b)} Spectrum of the toric code with $J=1,h=0.05$. The applied field breaks degeneracies in the energy levels and hybridizes states so that in each sector, every state has the same expectation value for the number of $e$'s and $m$'s present in the state.
    }
    \label{fig:eigenspectra}
\end{figure}

The energy spectrum for the toric code as computed with ED in the symmetry sector with $k_x=k_y=0$ is shown in Fig. \ref{fig:eigenspectra} with $h/J=0$ and $h/J=0.05$; two copies of each spectrum are shown with different color scales to show the values of $\langle N_e \rangle = (L_xL_y - \sum_v \langle A_v\rangle)/2$ and $\langle N_m \rangle = (L_xL_y - \sum_p \langle B_p\rangle)/2$ in each state. In the pure toric code case, the levels are flat and organized by the total number of excitations in each state; for example the level with $\langle N\rangle = \langle N_e\rangle +\langle N_m \rangle = 4$ features states with $(\langle N_e\rangle,\langle N_m\rangle)=(4e,0m)$, $(2e,2m)$, or $(0e,4m)$, each with an energy $4J$. Adding the magnetic field lifts the degeneracy in each level and hybridizes states, so that each state in the level has $\langle N_e\rangle = \langle N_m\rangle$; for example, in the $\langle N\rangle=4$ level, the states like $(0e,4m)$ mix with states like $(4e,0m)$ in order to form states like $(2e,2m)$.

\begin{figure*}
    \centering
    \includegraphics[width=0.95\linewidth]{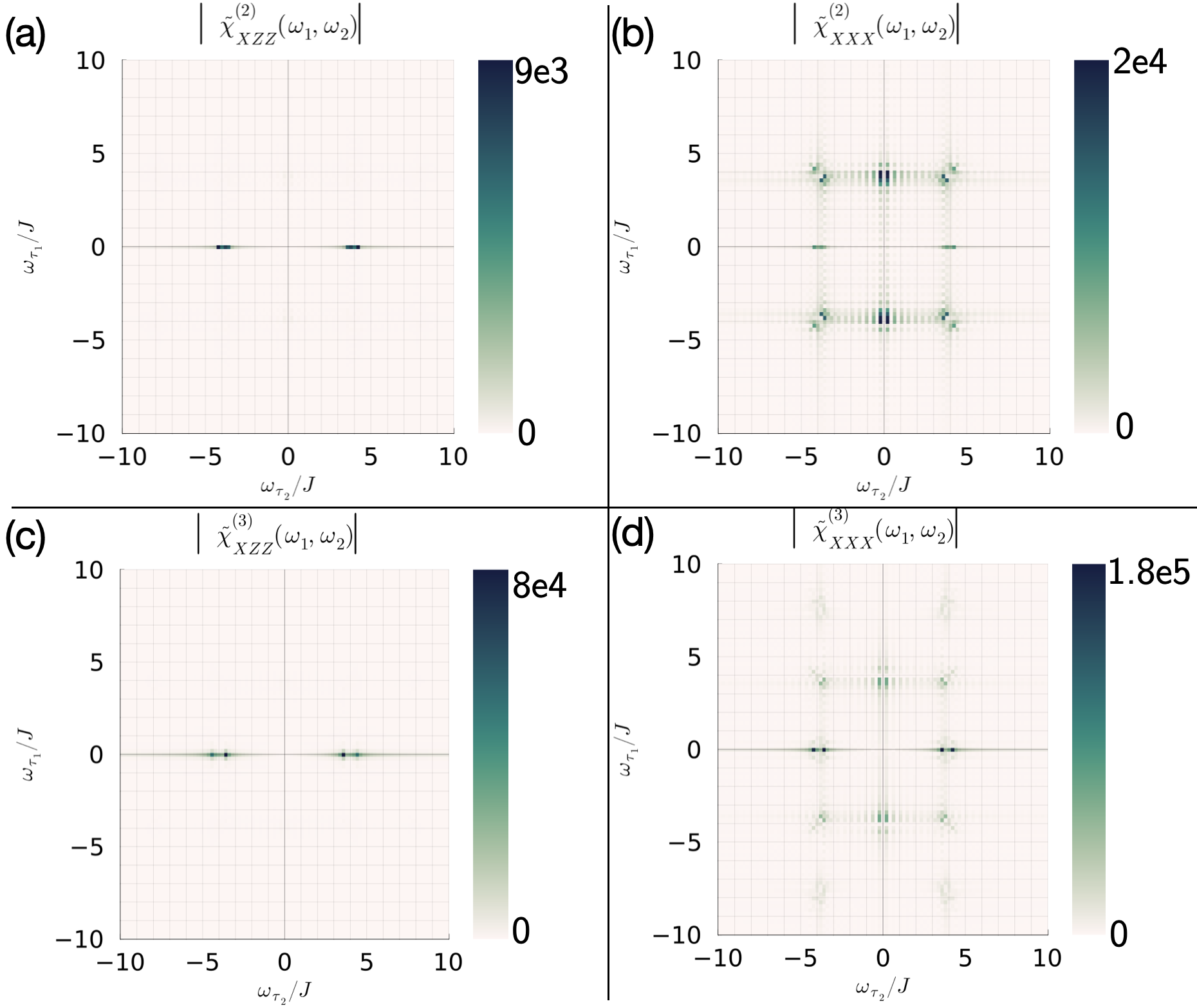}
    \caption{\textbf{Toric Code 2D Response Spectra:}
    2D response spectra plotted in the $\omega_{\tau_2} - \omega_{\tau_1}$ plane; specifically the magnitude of the two-dimensional Fourier transform  $\chi^{(n)}_{\alpha,\beta,\gamma}(\tau_1,\tau_2)$. Each of these is computed with ED on the $L_x\times L_y=3\times4$ toric code with $h/J=0.05$, up to $J\tau_1=J\tau_2=15$.
    \textbf{(a)} $\left|\tilde\chi^{(2)}_{XZZ}(\omega_{\tau_1},\omega_{\tau_2})\right|$
    \textbf{(b)} $\left|\tilde\chi^{(2)}_{XXX}(\omega_{\tau_1},\omega_{\tau_2})\right|$
    \textbf{(c)} $\left|\tilde\chi^{(3)}_{XZZ}(\omega_{\tau_1},\omega_{\tau_2})\right|$
    \textbf{(d)} $\left|\tilde\chi^{(3)}_{XXX}(\omega_{\tau_1},\omega_{\tau_2})\right|$
    }
    \label{fig:2dresponse-spectra}
\end{figure*}

As long as $h/J$ is kept small, we can rely on the effective QP picture and our prior analysis of allowed QP processes in Eqs. \ref{eq:chi2_definition} and \ref{eq:chi3_definition} to predict which frequencies we expect to see contributing in the response spectrum 
\begin{equation}
    \tilde\chi^{(n)}_{\alpha,\beta,\gamma}(\omega_{\tau_1},\omega_{\tau_2}) \equiv \mathcal{F}\left[\chi^{(n)}_{\alpha,\beta,\gamma}(\tau_1,\tau_2)\right](\omega_{\tau_1},\omega_{\tau_2})
    \label{eq:chi_tilde_defn}
\end{equation}
which is the two-dimensional Fourier transform of the time-domain data covered in the main text. In the case of $\chi^{(2)}_{XZZ}$, for a process like that shown in Fig. \ref{fig:chi2-plots}(a), we expect the signal to vanish due to the virtual decay of the excited $m$'s, but if we consider this decay as a background process to the dynamics of $e$'s carried out between $M^z(t_1)$ and $M^Z(t_2)$, then we would expect the signal to behave like $\propto e^{i2\Delta_e\tau_2}$, and thus we expect to a weak signal around $(\omega_{\tau_2},\omega_{\tau_1}) = (2\Delta_e,0)$ as confirmed by Fig. \ref{fig:2dresponse-spectra}(a). This approach has ignored the bandwidth of the anyon dispersion, so in reality the signal will be spread around $(2\Delta_e,0)$ with some structure dependent on the exact spectrum. (The other term in $\chi^{(2)}_{XZZ}$ behaves the same way, but with a phase of $-2\Delta_e$.) In $\chi^{(3)}_{XZZ}$, the first term acquires a phase $2\Delta_m\tau_1$ between $M^x(0)$ and $M^z(t_1)$, then a phase $2\Delta_m\tau_2 + 2\Delta_e\tau_2$ up until $M^z(t_2)$, but a phase of $-2\Delta_m(\tau_1+\tau_2)$ on the ket side back to $M^x(0)$, so the net result is a phase $2\Delta_e\tau_2$, and as such we see a signal around $(\omega_{\tau_2},\omega_{\tau_1})=(2\Delta_e,0)$ in Fig. \ref{fig:2dresponse-spectra}(c). 

As for $\chi^{(2)}_{XXX}$, the first term acquires a phase $2\Delta_m(\tau_1+\tau_2)$ between $M^x(0)$ and $M^x(t_2)$, then a phase $-2\Delta_m\tau_2$ between $M^x(t_2)$ and $M^x(t_1)$ for a net result of $2\Delta_m\tau_1$. The second term acquires $2\Delta_m\tau_1$ between $M^x(0)$ and $M^x(t_1)$, then $2\Delta_m\tau_2$ up to $M^x(t_2)$, for a total of $2\Delta_m(\tau_1+\tau_2)$. This explains the signals spotted around $(\omega_{\tau_2},\omega_{\tau_1})=(0,2\Delta_m)$ and $(2\Delta_m,2\Delta_m)$ in Fig. \ref{fig:2dresponse-spectra}(b). The relatively weaker signal around $(\pm2\Delta_m,0)$ can be attributed to the same type of processes creating the signal in $\tilde\chi^{(2)}_{XZZ}$, but with all $m$'s. The situation for $\chi^{(3)}_{XXX}$ admits many more possibilities in the cases where the operator $M^x(0)M^x(0)$ creates zero, two, or four $m$'s, which all combine to create signals at $(0,\pm2\Delta_m)$, $(\pm2\Delta_m, 0)$, $(\pm2\Delta_m,\pm2\Delta_m)$, and $(\pm2\Delta_m,\pm4\Delta_m)$, as seen in Fig. \ref{fig:2dresponse-spectra}(d).

We emphasize that none of the analysis in this Appendix indicated anything explicitly related to braiding; the signal observed in frequency space can be understood in terms of just the spectrum, its selection rules, and the dynamical phase acquired by different QP processes. The linear-in-$t$ early time growth behavior presented in the main text, however, explicitly reflects the growing probability of braiding occurring in the system.

\bibliography{references}

\end{document}